\documentclass[a4paper,11pt]{article}
\usepackage{jcappub} 
\usepackage{lineno,mathtools,bm}
\usepackage{ragged2e}
\usepackage[nameinlink,capitalise]{cleveref}
\usepackage{caption}
\usepackage{subcaption}
\usepackage{cancel}
\usepackage{soul}
\allowdisplaybreaks
\usepackage[dvipsnames]{xcolor}

\newcommand{\de}{{\rm d}}
\newcommand{\ka}{\bm {k}_1}
\newcommand{\kb}{\bm {k}_2}
\newcommand{\kc}{\bm {k}_3}

\newcommand{\ko}{\mathcal{K}}
\newcommand{\snr}{\ensuremath{{\rm SNR}}}

\let\Gamma\varGamma
\let\Delta\varDelta
\let\Theta\varTheta
\let\Xi\varXi
\let\Pi\varPi
\let\Sigma\varSigma
\let\Upsilon\varUpsilon
\let\Phi\varPhi
\let\Psi\varPsi
\let\Omega\varOmega

\def\sam#1{\textcolor{ForestGreen}{#1}}

\title{Decoupling Local Primordial non-Gaussianity from Relativistic Effects in the Galaxy Bispectrum}

\author[a,b]{Samantha J.\ Rossiter,}
\author[a,b,c,d]{Stefano Camera,}
\author[e,d]{Chris Clarkson,}
\author[d,f,g]{\\ Roy Maartens}

\affiliation[a]{Dipartimento di Fisica, Universit\`a degli Studi di Torino,\\
Via P.\ Giuria 1, 10125 Torino, Italy}
\affiliation[b]{INFN -- Istituto Nazionale di Fisica Nucleare, Sezione di Torino,\\
Via P.\ Giuria 1, 10125 Torino, Italy}
\affiliation[c]{INAF -- Istituto Nazionale di Astrofisica, Osservatorio Astrofisico di Torino,\\
Strada Osservatorio 20, 10025 Pino Torinese, Italy}
\affiliation[d]{Department of Physics \& Astronomy, University of the Western Cape,\\
Cape Town 7535, South Africa}
\affiliation[e]{Department of Physics \& Astronomy, Queen Mary University of London,\\
London E1 4NS, United Kingdom}
\affiliation[f]{Institute of Cosmology \& Gravitation, University of Portsmouth,\\
Portsmouth PO1 3FX, United Kingdom}
\affiliation[g]{National Institute for Theoretical and Computational Sciences,\\ 
Cape Town 7535, South Africa}

\emailAdd{samanthajosephine.rossiter@unito.it}

\abstract{Upcoming galaxy surveys aim to map the Universe with unprecedented precision, depth and sky coverage. The galaxy bispectrum is a prime source of information as it allows us to probe primordial non-Gaussianity (PNG), a key factor in differentiating various models of inflation. On the scales where local PNG is strongest,  Doppler and other relativistic effects become important and need to be included. Unlike for the single-tracer power spectrum, the leading order imaginary Doppler term does not cancel out in the bispectrum, leaving a smoking gun imaginary dipole signal. We investigate the detectability and importance of relativistic and local PNG contributions in the galaxy bispectrum. We compute the signal-to-noise ratio for the detection of lightcone projection effects in the bispectrum . Furthermore, we perform information matrix forecasts on the local PNG parameter, $f_{\rm NL}$, and on the parametrised amplitudes of the relativistic corrections. Finally, we quantify the bias on the measurement of $f_{\rm NL}$ that arises from neglecting relativistic effects. Our results show that detections of both first- and second-order relativistic effects are promising with forthcoming spectroscopic survey specifications -- and are largely unaffected by the uncertainty in $f_{\rm NL}$. Conversely, we show for the first time that neglecting relativistic corrections in the galaxy bispectrum can lead to a shift $>\!1.5\,\sigma(f_{\rm NL})$  on the detected value of $f_{\rm NL}$, highlighting the importance of including relativistic effects in our modelling.
}

\begin{document}
\maketitle
\flushbottom
\section{Introduction}
Advancements in observational cosmology, as well as sophisticated theoretical cosmological and statistical modeling, have propelled our understanding of the large-scale structure of the universe to new heights. With the wide variety of probes of large-scale structure, tracers of the underlying matter distribution, and their corresponding observational techniques and missions, the promise of new physics is riddled with nuance which requires detailed attention to the contributions that may distort, enhance, or cloud our visual description of the distribution of these structures.

Galaxy surveys have long been an indispensable probe of large-scale structure. We can infer the underlying physics that governs the evolution of matter perturbations by observing the galaxy distributions at different scales and redshifts, and therefore test the standard model of cosmology as well as theories of gravity. The anticipated measurements from current and future surveys such as the Dark Energy Spectroscopic Instrument (DESI) \cite{DESI:2016fyo,2023AJ....165..253H}, the \textit{Euclid} satellite mission \cite{Amendola:2016saw,2024arXiv240513491E}, the \textit{Nancy Grace Roman} Space Telescope \cite{2015arXiv150303757S,2021MNRAS.507.1746E} and MegaMapper \cite{Schlegel:2022vrv}, will provide information on ultra-large scales, i.e.\ beyond the matter-radiation equality scale. At these scales primordial non-Gaussianity (PNG) can be probed. It is also at these scales that effects from general relativity have the most impact on our measurements. This is the key motivation of this paper, quantifying how corrections to the galaxy number density which appear due to general relativity, specifically those derived from observing on our past lightcone, affect measurements of PNG. We explore this through the galaxy bispectrum, the Fourier transform of the 3-point correlation function of galaxy density contrasts, which is only non-zero in the presence of non-Gaussianity. Doppler shifts induced by the peculiar velocities of galaxies, gravitational lensing effects, and other relativistic corrections alter the clustering patterns of galaxies, imprinting distinct signatures on the galaxy distribution which can mimic those of PNG \cite{Bruni:2011ta,2012PhRvD..85b3504J,2015MNRAS.451L..80C, 2017JCAP...03..034U}. Therefore, understanding and accurately modeling these relativistic effects are imperative for extracting robust cosmological information from galaxy surveys. 

The $f_{\mathrm{NL}}$ parameter, which we use to quantify so-called local PNG, provides an excellent way to test theories of inflation, many of which depend heavily on constraints on $f_{\mathrm{NL}}$. PNG arises from departures of the primordial density fluctuations from a purely Gaussian distribution, a characteristic feature predicted by many inflationary models. Local PNG affects the bispectrum on ultra-large scales, which is also where the relativistic effects are strongest. Therefore, in a Newtonian bispectrum analysis, the unaccounted for relativistic contaminations would induce biases to the primordial signal, meaning a Gaussian primordial universe could be incorrectly found to be non-Gaussian \cite{Bruni:2011ta,2012PhRvD..85b3504J,2015MNRAS.451L..80C,Kehagias:2015tda,2017JCAP...03..034U,2017JCAP...09..040J, Koyama:2018ttg}. The best current constraints on $f_{\mathrm{NL}}$ are from Planck 2018 measurements of the CMB bispectrum:  $f_{\mathrm{NL}}=-0.9 \pm 5.1$ \cite{2020A&A...641A...9P}. Future CMB surveys will not be able to improve significantly on this, due to cosmic variance. In order to further improve $f_{\mathrm{NL}}$ precision, we need to access the much higher number of modes in galaxy surveys. Consequently, we need to fine-tune our modeling of relativistic corrections to the observables.

The galaxy power spectrum has been extensively studied and used to obtain impressive constraints on cosmological parameters. The dominant perturbation effect on sub-Hubble scales is from redshift-space distortions (RSD) \cite{1987MNRAS.227....1K}, which constitute the standard Newtonian approximation to projection effects. The inclusion of relativistic corrections to the galaxy density contrast at linear order \cite{2009PhRvD..80h3514Y,2011PhRvD..84f3505B,2011PhRvD..84d3516C} has, in turn, provided further insights. Many studies have been done on the possibility of constraining the local PNG parameter $f_{\mathrm{NL}}$ using the relativistic galaxy power spectrum for future galaxy and intensity mapping surveys, which avoids the bias typical in the Newtonian analysis \cite{Bruni:2011ta,2012PhRvD..85b3504J,2012PhRvD..85b3511L,2012arXiv1206.5809Y,Raccanelli:2013dza,Camera:2014bwa,Raccanelli:2015vla,Alonso:2015uua,Alonso:2015sfa,Fonseca:2015laa,Fonseca:2016xvi, Abramo:2017xnp, Lorenz:2017iez, Fonseca:2018hsu, Ballardini:2019wxj, Grimm:2020ays, Bernal:2020pwq,Wang:2020ibf}. However, the leading order relativistic correction to the observed galaxy overdensity is the imaginary Doppler term, which, for a single-tracer, becomes real in the power spectrum, and is thus harder to differentiate from the PNG effects they mimic. Whereas, in the bispectrum, these terms induce a smoking gun imaginary dipole signal \cite{Maartens:2019yhx}, making this higher order statistic a primary candidate for disentangling the two. PNG in the galaxy bispectrum has been extensively investigated in the Newtonian approximation \cite{Verde:1999ij,Scoccimarro:2003wn,Sefusatti:2006pa,Sefusatti:2007ih, 2010PhRvD..81f3530G,2011JCAP...04..006B,Tellarini:2015faa,Tellarini:2016sgp, Desjacques:2016bnm, Watkinson:2017zbs, Majumdar:2017tdm, Karagiannis:2018jdt,Yankelevich:2018uaz, Sarkar:2019ojl,Karagiannis:2019jjx, Bharadwaj:2020wkc, Karagiannis:2020dpq,MoradinezhadDizgah:2020whw}. In   \cite{Tellarini:2016sgp} it was shown that constraints on $f_{\mathrm{NL}}$ can be obtained from large-scale structure which are competitive with that of Planck, using forecasts on the Newtonian tree-level bispectrum (including RSD to second order), based on survey specifications from BOSS \cite{2013AJ....145...10D}, eBOSS \cite{eBOSS:2015jyv}, \textit{Euclid} \cite{2011arXiv1110.3193L} and DESI (LRGs, ELGs and QSOs) \cite{DESI:2013agm}.

To compute the relativistic galaxy bispectrum, we require the observed galaxy number density contrast to second order in perturbation theory, which has been done independently several times \cite{Bertacca:2014dra,Bertacca:2014wga,Yoo:2014sfa,DiDio:2014lka,Bertacca:2014hwa,Fuentes:2019nel,Magi:2022nfy}. At second order, the relativistic corrections are much more elaborate than those at first order because they involve quadratic couplings of first-order terms, introduce new terms such as the transverse peculiar velocity, lensing deflection angle and lensing shear, include second-order relativistic corrections to the galaxy bias, and require second-order gauge corrections to the second-order number density.

The relativistic corrections to the second order galaxy bias model have been calculated with Gaussian initial conditions \cite{Umeh:2019qyd} and in the presence of local PNG \cite{Umeh:2019jqg} (see \cite{Yoo:2022fun} for an alternative viewpoint). There are no such corrections to the first order galaxy bias, thus the tree-level power spectrum does not contain a relativistic correction to the bias model, but the tree-level bispectrum does. Furthermore, the local PNG signal in the tree-level galaxy power spectrum is sourced only by scale-dependent bias \cite{Dalal:2007cu,Matarrese:2008nc}, since there is no PNG signal in the primordial matter power spectrum at tree level. However, the local PNG in the galaxy bispectrum is sourced by scale-dependent bias, the primordial matter bispectrum, and RSD at second order \cite{Tellarini:2016sgp}. Notably, second-order relativistic corrections to RSD lead to new local PNG effects in the bispectrum. These effects arise from the coupling of first-order scale-dependent bias to first-order relativistic projection effects and the linearly evolved PNG in second-order velocity and metric potentials, which are absent in the standard Newtonian analysis.

In this paper, we compare current and future survey specifications by forecasting the amplitude of GR contributions, isolating the first-order and second-order corrections. We then perform an information matrix analysis to obtain the marginal errors on the $f_{\mathrm{NL}}$ parameter for local PNG and on each GR amplitude. We emphasise the importance of including GR contributions by calculating the bias on measurements of $f_{\mathrm{NL}}$ that is inherent in the Newtonian regime. Our work follows a series of papers which explore a particular model of the galaxy bispectrum \cite{2017JCAP...03..034U} that includes non-integrated, local lightcone projection effects \cite{Jolicoeur:2017eyi,2019JCAP...03..004J,Clarkson:2018dwn,Maartens:2019yhx,deWeerd:2019cae,Jolicoeur:2020eup, 2017JCAP...09..040J,  Umeh:2020cag}. We take exactly the model of the galaxy bispectrum from  \cite{Maartens:2020jzf}, which extends the previous list of works to incorporate local PNG into the relativistic bispectrum. This involves applying the previous results of \cite{Umeh:2019jqg,Umeh:2019qyd} on relativistic corrections to the second-order galaxy bias model. In line with the literature, we use the Fourier bispectrum, adopting a plane-parallel approximation.
\section{The galaxy bispectrum}\label{sec:GB}
We begin by outlining the model of the galaxy bispectrum for which we implement our analysis. The full description of this model can be found in \cite{Maartens:2020jzf}, which calculates the relativistic contributions to the galaxy bispectrum in the presence of local PNG by extending the derivation of the relativistic galaxy overdensity \cite{2017JCAP...03..034U,2017JCAP...09..040J,Jolicoeur:2017eyi, 2019JCAP...03..004J,Maartens:2019yhx}, neglecting integrated effects. Here we only present the key details relevant to our analysis. At tree level, the observed galaxy bispectrum is defined by
\begin{equation}
    \left<\Delta_{\rm g}^{(1)}(\bm{k}_1)\,\Delta_{\rm g}^{(1)}(\bm{k}_2)\,\Delta_{\rm g}^{(2)}(\bm{k}_3) \right> + 2\,\mathrm{c.p.} = 2 \, (2\pi)^3\,B_{\rm g}(\bm{k}_1,\bm{k}_2,\bm{k}_3) \,\delta^D(\bm{k}_1+ \bm{k}_2\,+ \bm{k}_3) \;,
\end{equation}
where $\mathrm{c.p.}$ denotes cyclic permutations over $\ka,\kb$, and $\kc$ and we use the convention that the observed density contrast is $\Delta_{\rm g} = \Delta_{\rm g}^{(1)} +\Delta_{\rm g}^{(2)}/2$.

The bispectrum is related at first order to the number density contrast at the source $\delta_{\rm g}$ by 
\begin{equation}
\Delta_{\rm g}=\delta_{\rm g}+\mathrm{RSD} + \text{relativistic effects} +\text{local PNG contributions}\;,
\end{equation}
and it can then be written in terms of first- and second-order Fourier kernels as
\begin{equation}
B_{\rm g}(\bm k_1, \bm k_2, \bm k_3)= \ko^{(1)}(\bm{k}_1)\,\ko^{(1)}(\bm{k}_2)\,\ko^{(2)}(\bm{k}_1,\bm{k}_2,\bm{k}_3)\,P(\ka)\,P(\kb) + 2\, \mathrm{c.p.}\;, 
\end{equation}
where $P$ is the linear matter power spectrum and the kernels $\ko^{(i)}$ are a summation of Newtonian, relativistic and local PNG contributions, i.e.\ $\ko^{(i)}=\ko^{(i)}_\mathrm{N} + \ko^{(i)}_\mathrm{GR} + \ko^{(i)}_\mathrm{PNG}$, for $i=1,2$. In writing the galaxy bispectrum this way, it is simple to isolate the effects we want to study. The Newtonian bispectrum is simply
\begin{equation}\label{eq:N bispectrum}
B_{\rm gN}(\bm k_1, \bm k_2, \bm k_3) =\ko_{\rm N}^{(1)}(\ka)\,\ko_{\rm N}^{(1)}(\kb)\,\ko_{\rm N}^{(2)}(\bm k_1, \bm k_2, \bm k_3)\,P(\ka)\,P(\kb)+2\,\mathrm{c.p.}\;,
\end{equation}
with
\begin{align}
\mathcal{K}_{\mathrm{N}}^{(1)}(\bm k_a)&=b_{10}+f\,\mu_a^2\;,\\
\mathcal{K}_{\mathrm{N}}^{(2)}(\bm k_1, \bm k_2, \bm k_3)&=b_{10}\,F_2(\bm k_1, \bm k_2)+b_{20}+f\,\mu_3^2\,G_2(\bm k_1, \bm k_2)+b_{s^2}\, S_2\,(\bm k_1, \bm k_2) \nonumber \\ 
& \quad+b_{10}\,f\,\left(\mu_1\,k_1+\mu_2\,k_2\right)\,\left(\frac{\mu_1}{k_1}+\frac{\mu_2}{k_2}\right)+f^2\, \frac{\mu_1\,\mu_2}{k_1\,k_2}\left(\mu_1\,k_1+\mu_2\,k_2\right)^2\;.
\label{kern2}
\end{align}
Here, $\mu_a=\hat{\bm k}_a \cdot \hat{\bm n}$, with $\hat{\bm n}$ the line-of-sight direction,  $f\coloneqq-\de\ln D / \de \ln(1+z)$ is the linear matter growth rate, and $D$ is the linear growth factor and $z$ the redshift. The kernels $F_2$ (mode-coupling part of the matter density contrast), $G_2$ (mode-coupling part of the matter velocity), and $S_2$ (tidal bias) are given in \cref{app:kernels}, with the linear and quadratic Gaussian clustering biases $b_{10}$ and $b_{20}$ given in \cref{app:g biases}. The second line of \cref{kern2} is the second-order RSD term. The contributions from local PNG or general relativity are omitted, apart from RSD -- which is such an essential effect to account for that it is now generally included in the basic Newtonian framework.

The local (non-integrated) general relativistic kernels at all orders of $\mathcal{H}/k$ are
\begin{align}
\mathcal{K}_{\mathrm{GR}}^{(1)}(\bm k_a)&=\mathrm{i}\,\mu_a\,\frac{\gamma_1}{k_a}+\frac{\gamma_2}{k_a^2}\;,
\label{eq:KGR1}
\\
\label{eq:KGR2}
\mathcal{K}_{\mathrm{GR}}^{(2)}(\bm k_1, \bm k_2, \bm k_3)&=\frac{1}{k_1^2\,k_2^2}\bigg\{ 
\,\beta_1+E_2(\bm k_1, \bm k_2, \bm k_3)\,\beta_2\,\nonumber \\
& \quad +\mathrm{i}\,\Big[\big(\mu_1\,k_1+\mu_2\,k_2\big)\,\beta_3+\mu_3\,k_3\,\big\{\beta_4+E_2(\bm k_1, \bm k_2, \bm k_3)\,\beta_5\big\}\Big] \nonumber\\
&\quad +\frac{k_1^2\,k_2^2}{k_3^2}\,\Big[F_2(\bm k_1, \bm k_2)\,\beta_6+G_2(\bm k_1, \bm k_2)\,\beta_7\Big]+\left(\mu_1\,k_1 \mu_2\,k_2\right)\,\beta_8 \nonumber\\
& \quad +\mu_3^2\,k_3^2\,\Big[\beta_9+E_2(\bm k_1, \bm k_2, \bm k_3)\,\beta_{10}\Big]+\left(\bm k_1 \cdot \bm k_2\right)\,\beta_{11}\nonumber \\
& \quad +\left(k_1^2+k_2^2\right)\,\beta_{12}+\left(\mu_1^2\,k_1^2+\mu_2^2\,k_2^2\right)\,\beta_{13}\nonumber \\
&\quad +\mathrm{i}\,\Big[\left(\mu_1\,k_1^3+\mu_2\,k_2^3\right)\,\beta_{14}+\left(\mu_1\,k_1+\mu_2\,k_2\right)\left(\bm k_1 \cdot \bm k_2\right)\,\beta_{15}\nonumber \\
&\quad  +k_1\,k_2\,\left(\mu_1\,k_2+\mu_2\,k_1\right)\,\beta_{16}+\left(\mu_1^3\,k_1^3+\mu_2^3\,k_2^3\right)\,\beta_{17} \nonumber\\
& \quad +\mu_1\,\mu_2\,k_1\,k_2\,\left(\mu_1\,k_1+\mu_2\,k_2\right)\,\beta_{18}+\mu_3\,\frac{k_1^2\,k_2^2}{k_3}\,G_2(\bm k_1, \bm k_2)\,\beta_{19}\Big]\bigg\}\;,
\end{align}
where 
\begin{align}\label{eq:gamma1}
\frac{\gamma_1}{\mathcal{H}} & =f\,\left[\mathcal E-2\,\mathcal{Q}-\frac{2\,(1-\mathcal{Q})}{\chi\,\mathcal{H}}-\frac{\mathcal{H}^{\prime}}{\mathcal{H}^2}\right]\;, \\
\frac{\gamma_2}{\mathcal{H}^2} & =f\,\left(3-\mathcal E\,\right)+\frac{3}{2}\,\Omega_{\rm m}\,\left[2+\mathcal E-f-4\,\mathcal{Q}-\frac{2\,(1-\mathcal{Q})}{\chi\,\mathcal{H}}-\frac{\mathcal{H}^{\prime}}{\mathcal{H}^2}\right]\;.\label{eq:gamma2}
\end{align}
In \cref{eq:KGR2}, $E_2$ is given in \cref{app:kernels} and the time-dependent beta functions $\beta_I$ are given in \cref{app:betas}. 
In \cref{eq:gamma1} and \cref{eq:gamma2}, $\chi$ is the radial comoving distance, $\mathcal{H}$ is the conformal Hubble rate, $\Omega_{\rm m}(z)=\Omega_{{\rm m}0}\,H_0^2\,(1+z)/\mathcal{H}(z)^2$ is the matter density at redshift $z$, and  two key astrophysical parameters in  relativistic analysis are introduced: the evolution bias $\mathcal E$ and the magnification bias $\mathcal Q$. These arise due to dependencies on the survey luminosity function (see \cref{app:MMB} for an example of how these biases are calculated).

Finally, the local PNG kernels are
\begin{align}
    \mathcal{K}_{\mathrm{nG}}^{(1)}(\bm k_a)&=\frac{b_{01}}{\mathcal{M}_a}~ \mbox{where} ~\mathcal{M}_a \equiv \mathcal{M}(k_a) \sam{=} \sqrt{\frac{25\,P(k_a)}{9\,P_{\cal R}(k_a)}} = \frac{10\,D}{3\,\mathcal{H}^2\,(3\,\Omega_{\rm m} +2\,f)} \, T(k_a)\,k_a^2 \;,\label{eq:Kng1}
    \\\mathcal{K}_{\mathrm{nG}}^{(2)}(\bm k_1, \bm k_2, \bm k_3)
    &=2\,f_{\mathrm{NL}}\,\left(b_{10}+f\,\mu_3^2\right)\,\frac{\mathcal{M}_3}{\mathcal{M}_1\,\mathcal{M}_2}+f\,b_{01}\,\left(\mu_1\,k_1+\mu_2\,k_2\right)\,\left(\frac{\mu_1}{k_1\,\mathcal{M}_2}+\frac{\mu_2}{k_2\,\mathcal{M}_1}\right) \nonumber \\
    & \quad +b_N\,N_2(\bm k_1, \bm k_2)+\frac{b_{11}}{2}\,\left(\frac{1}{\mathcal{M}_1}+\frac{1}{\mathcal{M}_2}\right)+\frac{b_{02}}{\mathcal{M}_1\,\mathcal{M}_2}\nonumber  \\
    & \quad +\frac{\mathcal{M}_3}{\mathcal{M}_1\,\mathcal{M}_2}\,\left(\frac{\Upsilon_1}{k_3^2}+\mathrm{i}\,\frac{\mu_3}{k_3}\,\Upsilon_2\right)+\Upsilon_3\,\left(\frac{1}{k_1^2\,\mathcal{M}_2}+\frac{1}{k_2^2\,\mathcal{M}_1}\right) \nonumber \\
    & \quad +\mathrm{i}\,\left[\Upsilon_4\,\left(\frac{\mu_1\,k_1}{k_2^2\,\mathcal{M}_1}+\frac{\mu_2\,k_2}{k_1^2\,\mathcal{M}_2}\right)+\Upsilon_5\,\left(\frac{\mu_1}{k_1\,\mathcal{M}_2}+\frac{\mu_2}{k_2\,\mathcal{M}_1}\right)\right] \;.\label{eq:Kng2}
\end{align}
In \cref{eq:Kng1}, $P_{\cal R}$ is the power spectrum of the primordial curvature perturbation, given by $P_{\cal R}(k)=2\,\pi^2\,k^{-3}\,A_{\rm s}\,(k/k_\ast)^{n_{\rm s}-1}$ with $A_{\rm s}$ the amplitude, $n_{\rm s}$ the spectral index, and $k_\ast$ the pivot wavenumber. The transfer function $T(k_a)$ relates the Gaussian potential to the linear primordial potential, i.e.\ $\varphi_{\text {in }}(\boldsymbol{k})=T(k)\,\varphi_{\mathrm{p}}(\boldsymbol{k})$ for $a_{\mathrm{p}} \ll a_{\mathrm{eq}} \ll a_{\mathrm{in}}$, where $a_{\mathrm{in}}$ is deep in the matter era.
The local PNG Fourier kernel, $N_2$, is given in \cref{app:kernels}, the time-dependent, $\Upsilon_I$, are given in \cref{app:upsilons} and the local PNG biases, $b_{01}$, $b_N$, $b_{11}$ and $b_{02}$, are given in \cref{nG biases}. 

It is important to note that the last two lines of \cref{eq:Kng2}, containing the $\Upsilon_I$ functions, arise from relativistic projection terms in $\Delta_g^{(2)}$ and thus count as both relativistic corrections and  local PNG contributions.
Therefore, the total galaxy bispectrum is made up of the Newtoninan galaxy bispectrum, the purely relativistic corrections, purely local PNG corrections, plus the cross terms. The Newtonian and RSD parts are real and have no overall $k$ scaling, while PNG contributions are real and contribute to even powers of ${\cal H}/k$. Relativistic Doppler contributions also contribute at odd powers of $H/k$. These odd-power terms are imaginary and contribute to the odd multipoles in general.

This model is the basis for our forecasting analysis. We expect there to be a degeneracy between PNG and relativistic contributions that scale as $({\cal H}/k)^2$, and our aim is to see if we can disentangle these. Because the relativistic parts also contribute at other powers of ${\cal H}/k$, we anticipate that we can. It is important to note that in our forecasts we omit some terms in the beta and upsilon functions, which are first order in ${\cal H}/k$, because of the lack of detailed information on the luminosity dependence of bias functions. We provide more insight into this caveat in \cref{sec:lum_derivs}.

\section{Methodology}
In this section, we outline the formalism employed for the signal-to-noise ratio (SNR), the information matrix and the bias on parameter estimation that arises from omitting local PNG and relativistic terms.\footnote{We shall at times refer to these as `shifts', when there might be ambiguity with respect to actual bias parameters.} We justify our choice of parameter space and survey specifications for optimization of these analyses. 
\label{sec:method}
\subsection{Signal-to-noise ratio}\label{sec:snr}
We extend the \snr\ analysis of \cite{Maartens:2019yhx} to our model with all orders of ${\cal H}/k$ and including local PNG. Under the assumption of Gaussian statistics for the bispectrum variance, leading to a diagonal covariance matrix, the \snr\ is given by 
\begin{equation}\label{eq:snr}
\snr^2(z)=\sum_{k_a, \mu_1, \varphi} \frac{1}{\operatorname{Var}\left[B_g\left(z, k_a, \mu_1, \varphi\right)\right]} B(z, k_a, \mu_1, \varphi)\,B^\ast(z, k_a, \mu_1, \varphi)\,D_B^2(z,k_a, \mu_1,\varphi)\;,
\end{equation}
with the (Gaussian) variance reading
\begin{equation}\label{eq:var}
\operatorname{Var}\left[B_g\left(z, k_a, \mu_1, \varphi\right)\right]=s_B\,\frac{\pi\,k_{\rm f}^3(z)}{k_1\,k_2\,k_3\,\Delta k^3} \frac{N_{\mu_1}}{\Delta \mu_1}\,\frac{N_{\varphi}}{\Delta\varphi}\,\tilde{P}_{\rm g \mathrm{N}}\left(z, k_1, \mu_1\right)\,\tilde{P}_{\rm g \mathrm{N}}\left(z, k_2, \mu_2\right)\,\tilde{P}_{\rm g \mathrm{N}}\left(z, k_3, \mu_3\right) .
\end{equation}
Note that the Newtonian approximation for the power spectrum is used in the variance, motivated by the fact that in the power spectrum of a single tracer, relativistic corrections enter at higher orders in \(\mathcal{H}/k\) relative to the bispectrum \cite{Maartens:2019yhx}. In \cref{eq:var}, 
\begin{equation}\label{eq:Ptilde}
\tilde{P}_{\rm g \mathrm{N}}\left(z, k_a, \mu_a\right)=P_{\rm g \mathrm{N}}\left(z, k_a, \mu_a\right)D_P(z,k_a)+\frac{1}{n_{\rm g}(z)}\;,
\end{equation}
where $P_{\rm g \mathrm{N}}$ the linear Newtonian galaxy power spectrum. In the expression for the variance, $s_B$ accounts for the triangle counting. Then \(k_{\rm f}\) is the fundamental frequency, determined by the comoving survey volume in the redshift bin under consideration, \(V_{\rm s}(z)\), through \(k_{\rm f}(z)=2\,\pi\,V_{\rm s}(z)^{-1/3}\). We choose the minimum wavenumber, $k_\mathrm{min}(z)$,  and the size of \(k\) bins, $\Delta k(z)$, as the fundamental frequency. Similarly, \(\Delta\mu_1\) and \(\Delta\varphi\) are the step lengths in the angular bins, and in our analysis we set $N_{\mu_1}=2$, $N_{\varphi}=2\,\pi$, $\Delta \mu_1=0.04$, $\Delta \varphi=\pi/25$. Finally, we adopt a redshift-dependent $k_\mathrm{max}=0.1\,(1+z)^{2 /\left(2+n_{\rm s}\right)}\,h\,\mathrm{Mpc}^{-1}$. In \cref{eq:snr,eq:Ptilde} we include a simple model of Finger-of-God (FoG) damping to account for non-linearities  due to RSD, which for the power spectrum and bispectrum respectively reads
\begin{align}
D_P(z,\bm k_a)&=\exp \left\{-\frac{1}{2}\,\left[k_a\,\mu_a\,\sigma(z)\right]^2\right\}\;,\\
D_B(z,\bm{k}_1,\bm{k}_2,\bm{k}_3)&= \exp \left\{-\frac{1}{2}\,\left[k_1^2\,\mu_1^2+k_2^2\,\mu_2^2+k_3^2\,\mu_3^2\right]\,\sigma(z)^2\right\}\;, 
\end{align}
where the $\sigma$ is the linear velocity dispersion. 

The $\snr(z)$ of \cref{eq:snr} refers to a single redshift bin centred at \(z\). Under the assumption of uncorrelated redshift bins, the total \snr\ is therefore simply given by the sum in quadrature of the $\snr(z_i)$ for each bin, \(i=1,\ldots, N_{\rm bins}\). Similarly, we can define a cumulative \snr\ by

\begin{equation}\label{eq:snrcum}
\snr(\leq z_i)^2={\sum_{j=1}^{i}\snr(z_j)^2}\;.
\end{equation}

\autoref{eq:snr}, as it is written, refers to the \snr\ for the overall detection of the bispectrum. 
Here, we are actually interested in assessing the possibility to measure relativistic and PNG contributions to the bispectrum. For this, it is sufficient to substitute \(B_{\rm g}\) in \cref{eq:snr} with other, appropriate expressions. Hence, we calculate the \snr\ for the following cases, in a fiducial cosmology with $f_{\rm NL}=0$ to signify no presence of local PNG. Specifically, we focus on:
\begin{itemize}
\item Total general relativistic contributions to the galaxy bispectrum,
    \begin{align}\label{eq:GRbisp}
    B_{\rm gGR}(\bm k_1, \bm k_2, \bm k_3) &= B_{\rm g}(\bm k_1, \bm k_2, \bm k_3)-B_{\rm gN} (\bm k_1, \bm k_2, \bm k_3)\nonumber \\
    &=\left[\ko^{(1)}(\ka)\,\ko^{(1)}(\kb)\,\ko^{(2)}(\bm k_1, \bm k_2, \bm k_3) \right. \nonumber \\
    &\left. \quad -\ko_{\rm N}^{(1)}(\ka)\,\ko_{\rm N}^{(1)}(\kb)\,\ko_{\rm N}^{(2)}(\bm k_1, \bm k_2, \bm k_3)\right]\,P(\ka)\,P(\kb)+2\,\mathrm{c.p.} \;.
    \end{align} 
\item Second-order general relativistic contributions to the galaxy bispectrum,
    \begin{align}\label{eq:GR2bisp}
    B_{\mathrm{gGR}}^{(2)}(\bm k_1, \bm k_2, \bm k_3)&=B_{\rm g}(\bm k_1, \bm k_2, \bm k_3) - \left[B_{\rm gN}(\bm k_1, \bm k_2, \bm k_3) + B_{\rm gGR}^{(1)}(\bm k_1, \bm k_2, \bm k_3)\right] \nonumber \\
    &= \Big\{\ko^{(1)}(\ka)\,\ko^{(1)}(\kb)\,\ko^{(2)}(\bm k_1, \bm k_2, \bm k_3) \nonumber \\
    &\quad -\left[\ko_{\rm N}^{(1)}(\ka)\,\ko_{\rm N}^{(1)}(\kb)\,\ko_{\rm N}^{(2)}(\bm k_1, \bm k_2, \bm k_3) \right. \nonumber \\
    & \quad+ \left. \left( \ko_{\rm N}^{(1)}(\ka)\,\ko_{\rm GR}^{(1)}(\kb)\,+\ko_{\rm GR}^{(1)}(\ka)\,\ko_{\rm N}^{(1)}(\kb) \right.\right.\nonumber\\
    &\quad+ \left. \left.\ko_{\rm GR}^{(1)}(\ka)\,\ko_{\rm GR}^{(1)}(\kb)\,\right)\ko_{\rm N}^{(2)}(\bm k_1, \bm k_2, \bm k_3) \right]\Big\}\,P(\ka)\,P(\kb) \nonumber \\
    &\quad +2\,\mathrm{c.p.}\;.
    \end{align}
\end{itemize}
Note that in both cases, we assume as fiducial \(f_{\rm NL}=0\).
\subsection{Information matrix}\label{sec:Fisher}
In order to quantify the amount of information we can obtain from this model of the galaxy bispectrum, we employ a information matrix formalism. Assuming the same notation as \cref{eq:snr} and bispectrum estimator variance as in  \cref{eq:var}, we have a very simple form for the information matrix for the bispectrum in the case of a Gaussian likelihood, namely
\begin{equation}
    F_{\alpha\beta}=\sum_{z,k_a,\mu_a,\varphi}\frac{\partial_{(\alpha}B_{\rm g}\,\partial_{\beta)}B_{\rm g}^\ast}{{\rm Var}[B_{\rm g},B_{\rm g}]}\;,\label{eq:Fisher}
\end{equation}
where $\partial_\alpha=\partial/\partial\theta_\alpha$ and $\theta_\alpha$ are the parameters under consideration. Round brackets denote symmetrisation. As with the SNR, we sum over orientation bins $\mu_a$ and $\varphi$ and wave-mode bins $k_a$, and then over redshift bins $z$. In the Gaussian likelihood approximation, this is the inverse of the covariance matrix, so we obtain the marginal errors on our parameters by
\begin{equation}\label{eq:marginal error}
    \sigma (\theta_\alpha)=\sqrt{\left({\sf F}^{-1}\right)_{\alpha \alpha}}\;.
\end{equation}

\subsection{Parameter space}\label{sec:paramspace}
Local PNG contributions are parametrised by  $f_{\mathrm{NL}}$. In order to have similar parameters for the relativistic corrections, we introduce arbitrary amplitudes in the relevant kernels. Explicitly,
\begin{align}\label{eq:B_GR}
B_{\rm g}(\bm k_1, \bm k_2, \bm k_3) &= \left[N^{(1)}(\ka,\kb)\, N^{(2)}(\bm k_1, \bm k_2, \bm k_3)+ \epsilon^{(1)}_{\rm GR}\, N^{(2)}(\bm k_1, \bm k_2, \bm k_3) \,GR^{(1)}( \ka,\kb) \right. \nonumber \\
& \quad \left. + \epsilon^{(1)}_{\rm GR}\, \epsilon^{(2)}_{\rm GR} \,GR^{(1)}(\ka,\kb)\,GR^{(2)}(\bm k_1, \bm k_2, \bm k_3) \right. \nonumber \\
&\quad \left. + \epsilon^{(2)}_{\rm GR}\,GR^{(2)}(\bm k_1, \bm k_2, \bm k_3)\, N^{(1)}(\ka,\kb) \right]\,P(\ka)\,P(\kb)+2\,\mathrm{c.p.} \; ,
\end{align}
where
\begin{align} \label{eq:subs}
N^{(1)}(\ka,\kb) &= \ko^{(1)}_{\rm N}(\ka)\,\ko^{(1)}_{\rm N}(\kb)\,+\ko^{(1)}_{\rm N}(\ka)\,\ko^{(1)}_{\rm nG}(\kb)\, \nonumber \\ 
&\quad + \ko^{(1)}_{\rm nG}(\ka)\,\ko^{(1)}_{\rm N}(\kb)\,+ \ko^{(1)}_{\rm nG}(\ka)\,\ko^{(1)}_{\rm nG}(\kb) \; ,\nonumber  \\
GR^{(1)}(\ka,\kb)&=\ko^{(1)}_{\rm N}(\ka)\,\ko^{(1)}_{\rm GR}(\kb)\,+\ko^{(1)}_{\rm GR}(\ka)\,\ko^{(1)}_{\rm N}(\kb)\,+\ko^{(1)}_{\rm GR}(\ka)\,\ko^{(1)}_{\rm GR}(\kb)\, \nonumber \\
&\quad +\ko^{(1)}_{\rm GR}(\ka)\,\ko^{(1)}_{\mathrm{nG}}(\kb)\,+\ko^{(1)}_{\mathrm{nG}}(\ka)\,\ko^{(1)}_{\rm GR}(\kb) \; ,\nonumber  \\
N^{(2)}(\bm k_1, \bm k_2, \bm k_3)&= \ko^{(2)}_\mathrm{N}\,(\bm k_1, \bm k_2, \bm k_3) +\ko^{(2)}_\mathrm{nGnewt}\,(\bm k_1, \bm k_2, \bm k_3)  \; , \nonumber \\
GR^{(2)}(\bm k_1, \bm k_2, \bm k_3)&= \ko^{(2)}_\mathrm{GR}\,(\bm k_1, \bm k_2, \bm k_3) +\ko^{(2)}_\mathrm{nGgr}\,(\bm k_1, \bm k_2, \bm k_3) \; .
\end{align}
 Here we have a first-order $\epsilon_{\mathrm{GR}}^{(1)}$ and a second-order $\epsilon_{\mathrm{GR}}^{(2)}$ relativistic amplitude and all $\ko_{\mathrm{nG}}$ are dependent on $f_{\mathrm{NL}}$. $\ko^{(2)}_\mathrm{nGnewt}\,\mathrm{and}\,\ko^{(2)}_\mathrm{nGgr}$ denote the Newtonian and GR parts of \cref{eq:Kng2} respectively. We can now state our chosen parameter space as 
\begin{align*}
 \{\theta_\alpha\} = \Big\{\epsilon^{(1)}_{\rm GR},\epsilon^{(2)}_{\rm GR},f_{\mathrm{NL}}\Big\}\;.   
\end{align*}

\subsection{Bias on parameter estimation} \label{sec:param bias}
We quantify the importance of including local primordial non-Gaussianity and relativistic contributions in our analysis by following \citep[][see also \citealp{2015MNRAS.451L..80C}]{2007MNRAS.380.1029H} on shifts (also referred to as `bias') on parameter estimation. The shift in the best-fit value of a parameter due to an incorrect assumption on the value of $f_{\rm NL}$, is given by
\begin{equation}
\Delta \theta_\alpha=\sum_\beta\left(F^{\theta_{\rm GR} \theta_{\rm GR}}\right)_{\alpha \beta}^{-1} F_\beta^{\theta f_{\rm NL}}\,\delta f_{\rm NL} \;.
\end{equation}
Here $\theta_{\rm GR}$ is our chosen parameterisation of the GR contributions and $\delta f_{\rm NL}=|f_{\rm NL}^{\rm(true)}-f_{\rm NL}^{\rm(ref)}|$ is the amount by which we are wrong in our assumption about the reference value of \(f_{\rm NL}\), relative to its true value. Since we always assume the absence of primordial non-Gaussianity, i.e.\ $f_{\rm NL}^{\rm(true)}=0$, it follows that $\delta f_{\rm NL}$ equals the fiducial value assumed in the analysis. Applying this to the case where $\theta_{\rm GR} = \{\epsilon_{\rm GR}^{(1)},\epsilon_{\rm GR}^{(2)}\}$ gives 
\begin{align}\label{eq:grbias}
\Delta \theta_{\epsilon_{\rm GR}^{(1)}} &= \left[\left(F^{\theta_{\rm GR} \theta_{\rm GR}}\right)_{\epsilon_{\rm GR}^{(1)} \epsilon_{\rm GR}^{(1)}}^{-1} F_{\epsilon_{\rm GR}^{(1)} f_{\rm NL}} + \left(F^{\theta_{\rm GR} \theta_{\rm GR}}\right)_{\epsilon_{\rm GR}^{(1)} \epsilon_{\rm GR}^{(2)}}^{-1} F_{\epsilon_{\rm GR}^{(2)} f_{\rm NL}}\right]\,\delta f_{\rm NL}
\; , \nonumber \\
\Delta \theta_{\epsilon_{\rm GR}^{(2)}} &= \left[\left(F^{\theta_{\rm GR} \theta_{\rm GR}}\right)_{\epsilon_{\rm GR}^{(2)} \epsilon_{\rm GR}^{(1)}}^{-1} F_{\epsilon_{\rm GR}^{(1)} f_{\rm NL}} 
+ \left(F^{\theta_{\rm GR} \theta_{\rm GR}}\right)_{\epsilon_{\rm GR}^{(2)} \epsilon_{\rm GR}^{(2)}}^{-1} F_{\epsilon_{\rm GR}^{(2)} f_{\rm NL}}\right] \,\delta f_{\rm NL} \;.
\end{align}

On the other hand, to quantify the shift on a measurement of $f_{\rm NL}$ due to neglecting GR effects in the modelling of our signal, we have
\begin{equation}\label{eq:fnlbias}
    \Delta \theta_{f_{\rm NL}} =\left(F_{{f_{\rm NL}} {f_{\rm NL}}}\right)^{-1} \left(F_{f_{\rm NL}\epsilon_{\rm GR}^{(1)}} +  F_{f_{\rm NL}\epsilon_{\rm GR}^{(2)}}\right)\;.
\end{equation}
Note that, since \(\epsilon_{\rm GR}^{(1)}\) and \(\epsilon_{\rm GR}^{(2)}\) are binary amplitude parameters, corresponding to no GR effects if they are set to zero, their corresponding \(\delta\epsilon_{\rm GR}^{(1)}\) and \(\delta\epsilon_{\rm GR}^{(2)}\) are equal to 1 by construction.
\section{Results} \label{sec:results}
\begin{figure}[h!]
\centering
\includegraphics[width=.45\textwidth]{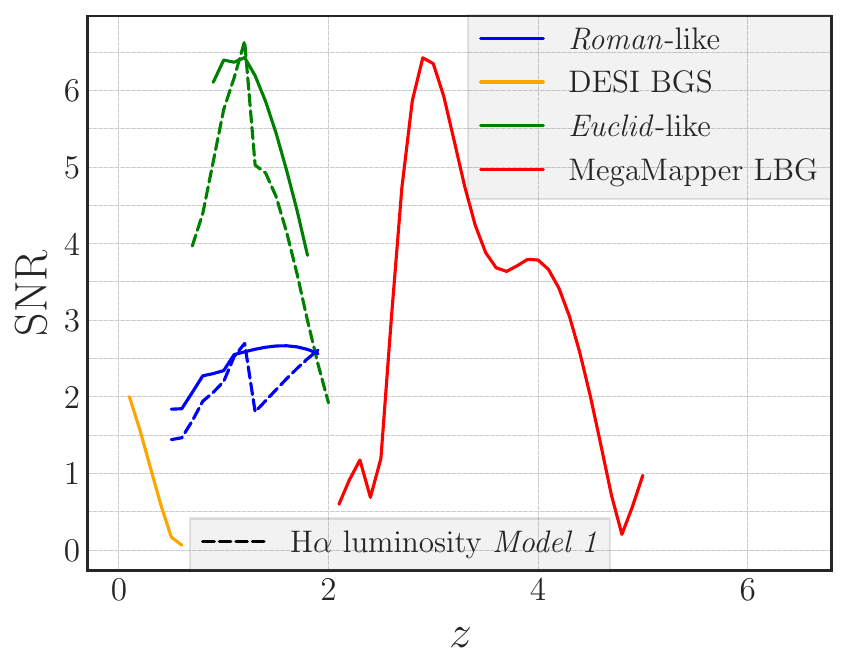}
\qquad
\includegraphics[width=.45\textwidth]{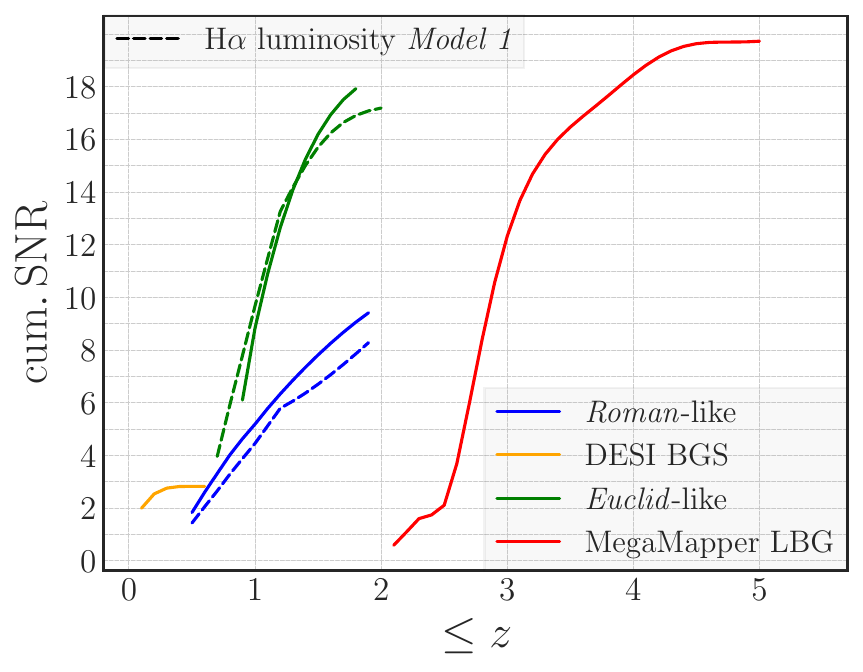}
\qquad
\caption{Signal-to-noise ratio for the general relativistic contributions to the galaxy bispectrum for each survey. The dashed blue and green  lines represent the \textit{Model 1} parameters for the $\mathrm{H}\alpha$ surveys, \textit{Roman}- and \textit{Euclid}-like respectively.}
\label{fig:surv_snrGR}
\end{figure}

Following \cref{sec:method} and its  subsections on the methodology, we compare the results of the information matrix analysis outlined in \cref{sec:Fisher} applied to three benchmark surveys: a low-redshift DESI-like bright galaxy survey (BGS), two space-borne, mid-to-high redshift surveys like \textit{Roman} and \textit{Euclid}, and a more futuristic, high-redshift spectroscopic galaxy survey like MegaMapper (targetting Lyman-break galaxies, `LBGs'). Moreover, for the two satellites, we consider two different modelling of their source population, dubbed (according to the literature), \textit{Model 1} and \textit{Model 3}. The analysis is done for $f_{\mathrm{NL}}=0$, firstly comparing the survey SNRs, using the formalism set out in \cref{sec:snr} for the total general relativistic contribution (\cref{fig:surv_snrGR}) and the second-order general relativistic contribution (\cref{fig:surv_snrGR2}).
\subsection{Galaxy survey specifications} \label{sec:surv specs}
The focus of our forecast is on current and future spectroscopic galaxy surveys, both ground-based and space-borne. For the former category, we implement specifications for DESI's Bright Galaxy Sample (BGS) and MegaMapper LBG, a future wide field spectroscopic instrument currently under development, for which we use the Lyman-Break Galaxy (LBG) target specifications. For the latter, we consider a H$\alpha$ emission-line galaxy population, as will be detected by the spectroscopic survey from the Wide Field Instrument on board the \textit{Nancy Grace Roman} Space Telescope (\textit{Roman}, hereafter) and the Near Infrared Spectroscopic and Photometric instrument equipped by the European Space Agency's \textit{Euclid} space satellite. The difference in scope of these surveys can be seen in \cref{tab:surveys}. For those surveying H$\alpha$ emitters, we consider two cases, using survey parameters drawn from so-called \textit{Model 1} and \textit{Model 3} luminosity functions, as first introduced by \cite{Pozzetti:2016cch}. Previous analysis of the SNR of the Doppler contribution to the galaxy bispectrum for a \textit{Euclid}-like survey has used a Schechter-type luminosity function, \textit{Model 1}. This model yields an SNR  $\mathcal{O}(10)$ for the leading order Doppler bispectrum \cite{Maartens:2019yhx}. However the discontinuity of this luminosity function can be seen in the SNR and is an artefact of the model, having no physical significance. We therefore adapt the \textit{Model 1} analysis to the galaxy bispectrum model outlined in \cref{sec:GB}, isolating the general relativistic bispectrum as in \cref{eq:B_GR},  in order to compare the SNR for luminosity \textit{Model 3} \citep{Maartens:2021dqy}, which is derived by fitting from simulations, with the updated \textit{Euclid} redshift range.

\begin{figure}[htbp]
\centering
\includegraphics[width=.4\textwidth]{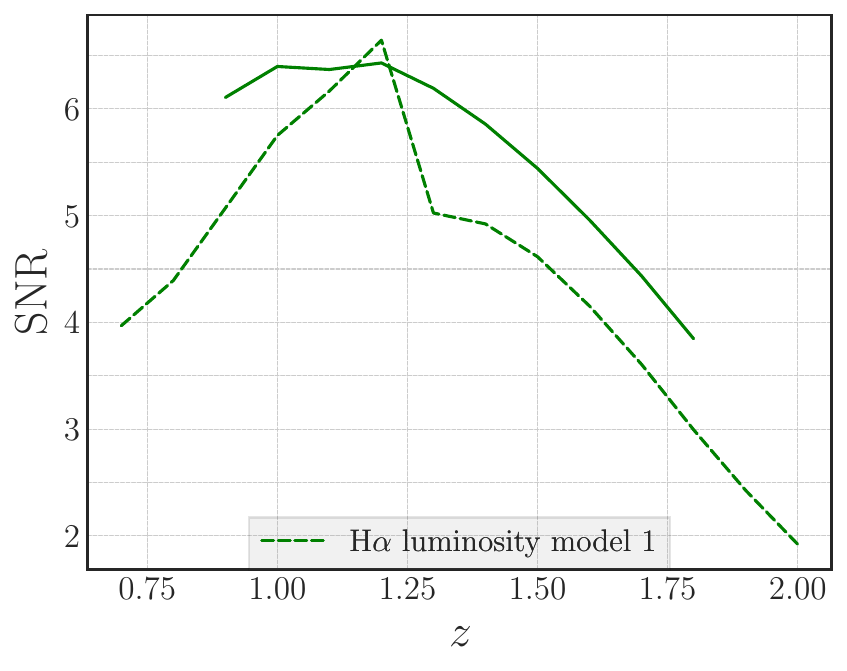}
\qquad
\includegraphics[width=.4\textwidth]{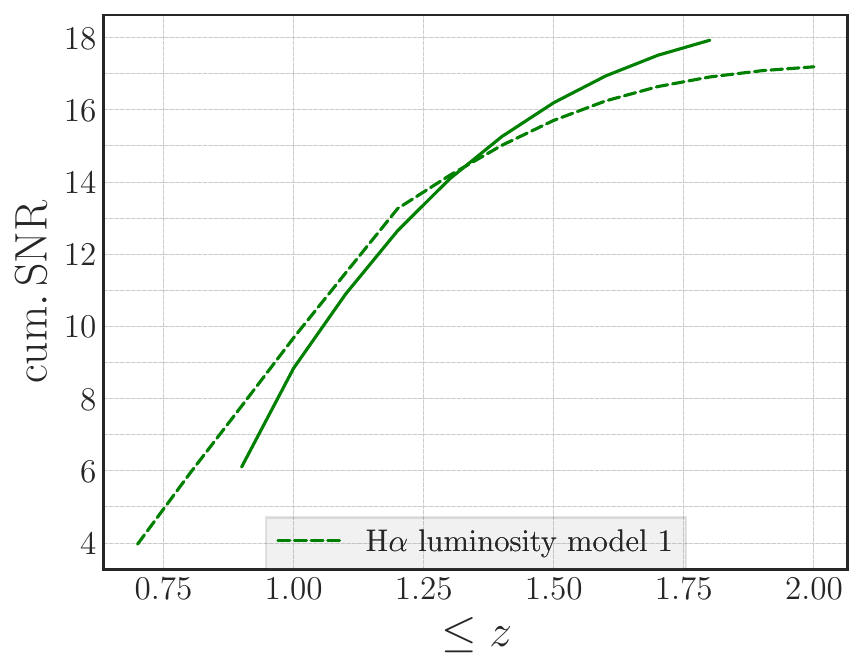}
\caption{The signal-to-noise ratio (left: per redshift bin, right: cumulative) of the general relativistic contribution to the bispectrum for a \textit{Euclid}-like survey with luminosity \textit{Model 1} (dashed) and \textit{Model 3} (solid). 
Local primordial non-Gaussianity is neglected ($f_{\mathrm{NL}}=0).$\label{fig:SNRmodels}}
\end{figure}

The results are displayed in \cref{fig:SNRmodels}, which interestingly show that even with a smaller redshift range, \textit{Model 3} achieves a slightly better SNR than  \textit{Model 1}. The full information matrix forecast results for both models applied to \textit{Roman}- and \textit{Euclid}-like, the two surveys targeting H$\alpha$ emitters, are given in \cref{sec:results}.

\begin{table}[htbp]
\centering
\caption{Specifications for the adopted mock surveys.\label{tab:surveys}}
\begin{tabular}{c|c|c}
\hline
Survey & Redshift range& Sky fraction ($f_{\mathrm{sky}}$)\\
\hline
\textit{Roman}-like & \textit{Model 1 \& 3:} $0.5\leq z \leq 1.9$ & $0.05$ \\
\hline
DESI BGS& $0\leq z \leq 0.6$  & $0.34$  \\
\hline
\textit{Euclid}-like & \begin{tabular}{c}\textit{Model 1}: $0.7\leq z \leq 2$ \\ \textit{Model 3}: $0.9\leq z \leq 1.8$\end{tabular}  & $0.36$ \\
\hline
MegaMapper LBG& $2.1\leq z \leq 5$ & $0.50$ \\
\hline
\end{tabular}
\end{table}
For our analysis we use cosmological parameters from Planck 2018 data \cite{Planck:2018vyg}: $h = 0.6766$, $\Omega_{\rm m0} =0.3111$, $\Omega_{\rm b0}\,h^2 = 0.02242$, $\Omega_{\rm c0}\,h^2 = 0.11933$, $n_{\rm s} = 0.9665$, $\sigma_8= 0.8102$, $\gamma = 0.545$, $\Omega_{K0} = 0 = \Omega_{\nu0}$. The comoving survey volume is given by
\begin{equation}
    V(z)=\frac{4}{3}\,\pi\,f_{\rm sky}\,\left[\chi^3(z+\Delta z /2) - \chi^3(z-\Delta z/2)\right]\;,
\end{equation}
with \(f_{\rm sky}\) the fraction of surveyed sky and \(\Delta z=0.1\) is the width of the redshift bin which we have used for all surveys. The galaxy number density $n_{\rm g}$, magnification bias $\mathcal{Q}$ and evolution bias $\mathcal{E}$ however are dependent on the luminosity function specific to the survey. The explicit implementations for deriving these values for our H$\alpha$ surveys (\textit{Euclid}/\textit{Roman}-like) and K-corrected survey (DESI BGS) are outlined in \citep{Maartens:2021dqy}. We obtained the MegaMapper LBG values by taking the Schechter function form for the UV luminosity function \cite{Sailer:2021yzm} and best-fit parameters from table 3 of \cite{Wilson:2019brt}. The survey-specific input parameters are given in this paper in \cref{app:sp}.

\begin{figure}[h!]
\centering
\includegraphics[width=.45\textwidth]{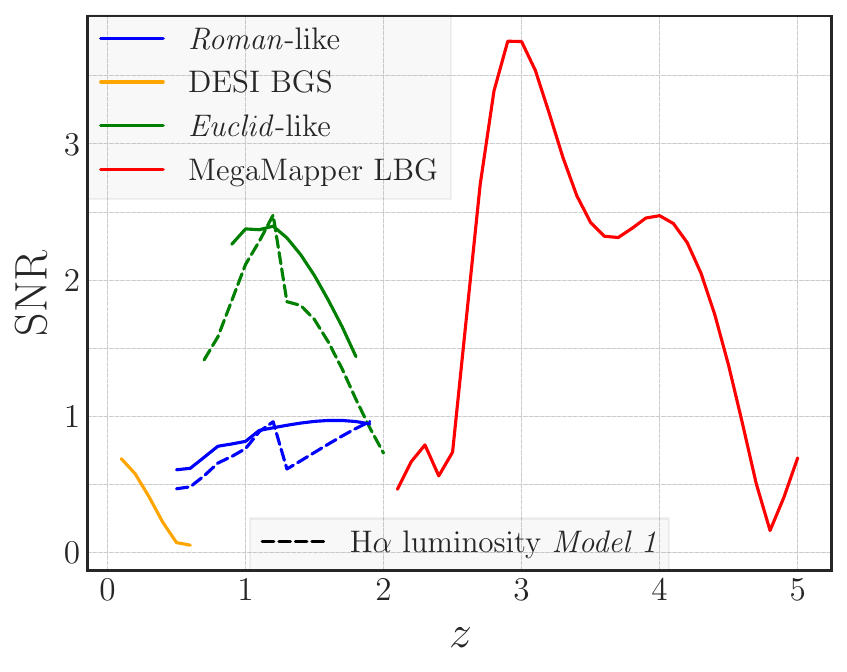}
\qquad
\includegraphics[width=.45\textwidth]{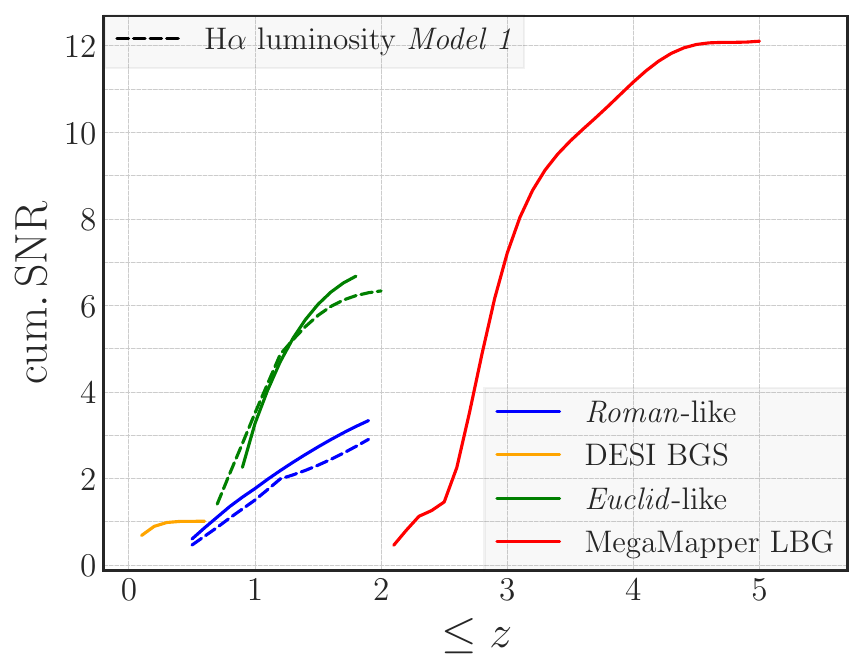}
\qquad
\caption{Same as \cref{fig:surv_snrGR} but for only the second-order GR corrections.}
\label{fig:surv_snrGR2}
\end{figure}

From these plots we can clearly see the promise of a detectable GR signal with upcoming surveys, especially MegaMapper LBG and a \textit{Euclid}-like survey,
but not with DESI BGS. We obtain a total GR signal of $\sim 19$ for MegaMapper LBG, only slightly higher than for a \textit{Euclid}-like survey. Naively, we would expect a larger signal from MegaMapper LBG considering its larger redshift range and therefore ability to probe small $k$ well beyond the equality scale. However our analysis shows that the signal is significantly suppressed by the evolution bias, $\mathcal E$. We find for the second-order GR bispectrum an expected  total SNR of $\sim 6$ with a \textit{Euclid}-like survey and an impressive SNR of $\sim 11$ for MegaMapper LBG, as seen in \cref{fig:surv_snrGR2}, an improvement which supports our initial expectations of the capabilities of the MegaMapper project. It is interesting to note that the SNR for $\mathrm{H}\alpha$ surveys, \textit{Euclid} and \textit{Roman}-like, using luminosity \textit{Model 3} is larger than for luminosity \textit{Model 1} in both examples, even with a smaller redshift range in the case of a \textit{Euclid}-like survey. 

\begin{figure}[t]
    \centering
    \begin{minipage}[t]{0.49\textwidth}
        \centering
        \includegraphics[width=\textwidth]{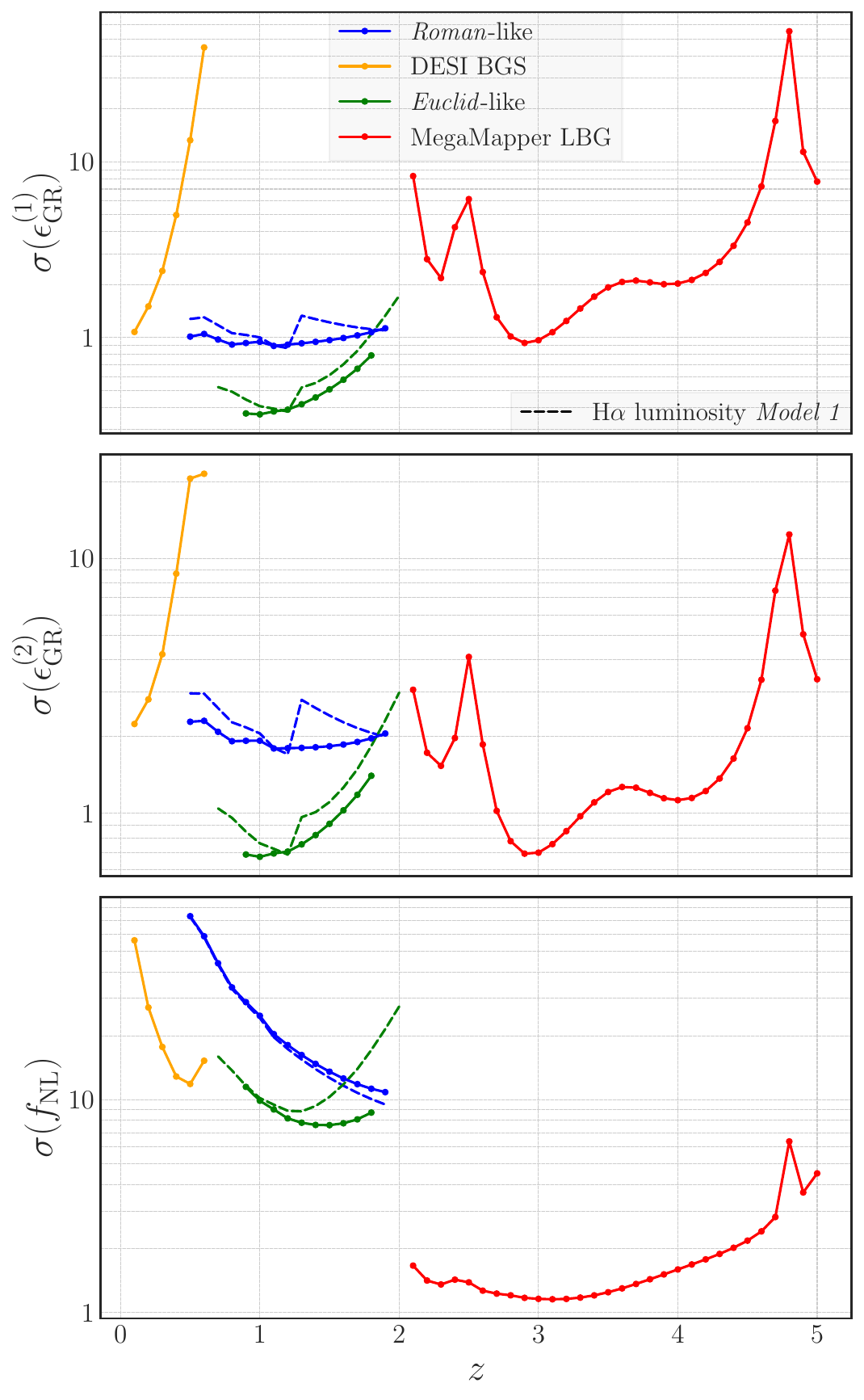}
    \end{minipage}
    \hfill
    \begin{minipage}[t]{0.49\textwidth}
        \centering
        \includegraphics[width=\textwidth]{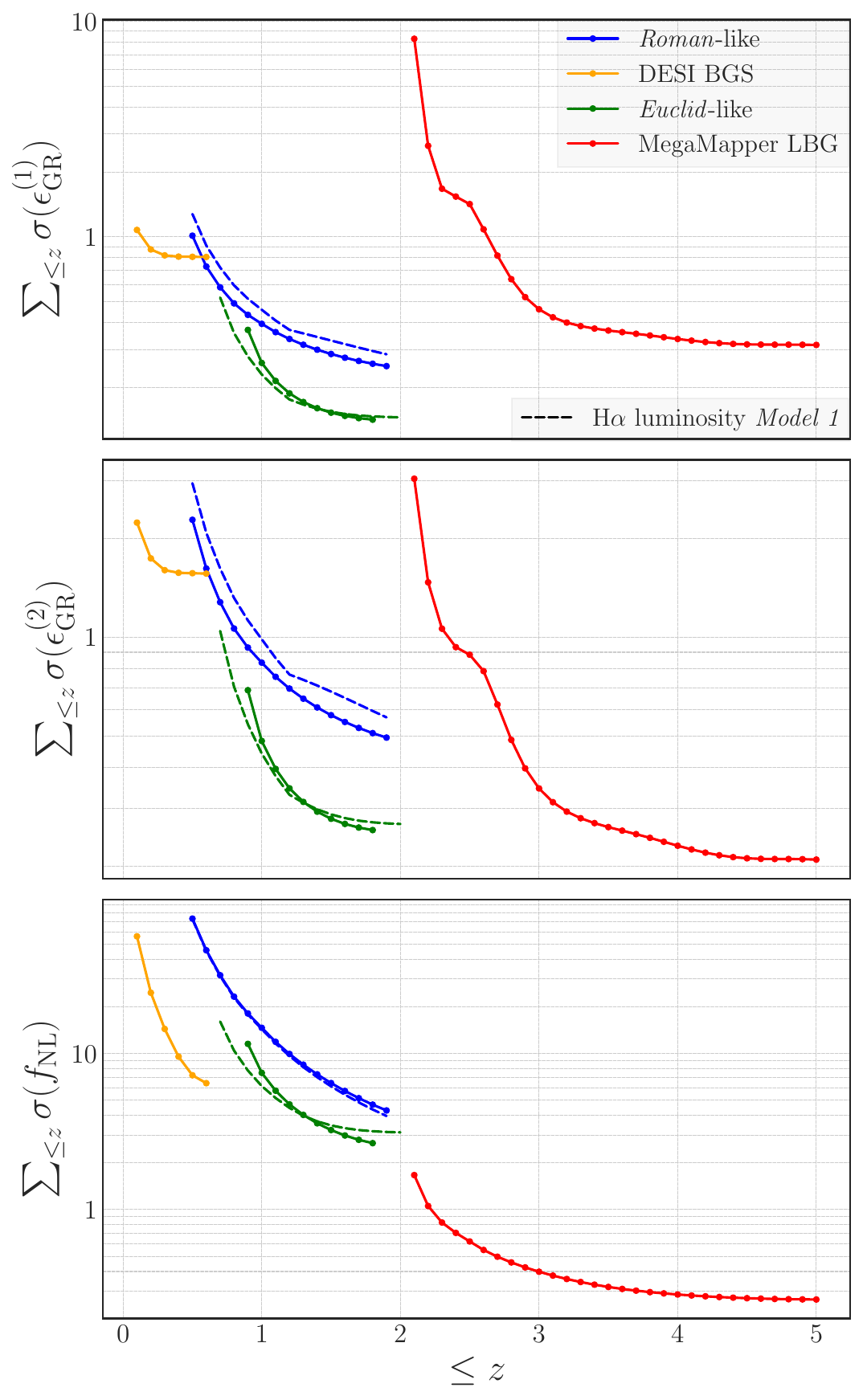}
    \end{minipage}
    \caption{Marginal errors (left: per redshift bin, right: cumulative) on the parameters $\epsilon_{\mathrm{GR}}^{(1)}$, $\epsilon_{\mathrm{GR}}^{(2)}$, and $f_{\mathrm{NL}}$ for $f_{\mathrm{NL}}=0$.}
    \label{fig:comperrors}
\end{figure}

This is largely in agreement with the results of the Fisher information matrix analysis on the parameter space $\theta=\{\epsilon_{\rm GR}^{(1)}, \epsilon_{\mathrm{GR}}^{(2)}, f_{\mathrm{NL}}\}$ as outlined in \cref{sec:Fisher} and \cref{sec:paramspace}. The marginal errors are given in \cref{fig:comperrors}, showing the comparatively impressive capacity of a \textit{Euclid}-like survey in detecting the general relativistic signal at both orders, achieving a total marginal error $\sigma (\epsilon_{\mathrm{GR}}^{(1)})<0.15$ and $\sigma (\epsilon_{\mathrm{GR}}^{(2)})<0.27$. Our results show that MegaMapper LBG would not improve on the precision of the total relativistic detection, being comparable with a \textit{Roman}-like survey, even with the vastly larger survey scope, as predicted by the SNR. It may however provide further information on the second-order GR corrections, since we find a total marginal error of $\sim 0.22$ for $\epsilon_{\mathrm{GR}}$.

\begin{table}[h!]
\centering
\caption{Marginal errors on the parameters $\epsilon_{\rm GR}^{(1)}, \epsilon_{\mathrm{GR}}^{(2)}, f_{\mathrm{NL}}$ for each adopted survey.\label{tab:errors}}
\begin{tabular}{lccc}
\hline
Survey & $\sigma (\epsilon_{\mathrm{GR}}^{(1)})$&  $\sigma (\epsilon_{\mathrm{GR}}^{(2)})$& $\sigma (f_{\mathrm{NL}})$\\
\hline
\hline
\textit{Roman}-like \textit{Model 1} & 0.29 & 0.57 & 4.00 \\
\hline
\textit{Roman}-like \textit{Model 3} & 0.25 & 0.50 & 4.30 \\
\hline
DESI BGS& 0.81 & 1.60 & 6.40  \\
\hline
\textit{Euclid}-like \textit{Model 1}& 0.15 & 0.27 & 3.10 \\
\hline
\textit{Euclid}-like \textit{Model 3}& 0.14 & 0.26 & 2.70 \\
\hline
MegaMapper LBG & 0.32 & 0.21 &0.26\\
\hline
\end{tabular}
\end{table}

\begin{figure}[h!]
\centering
\includegraphics[width=.6\textwidth]{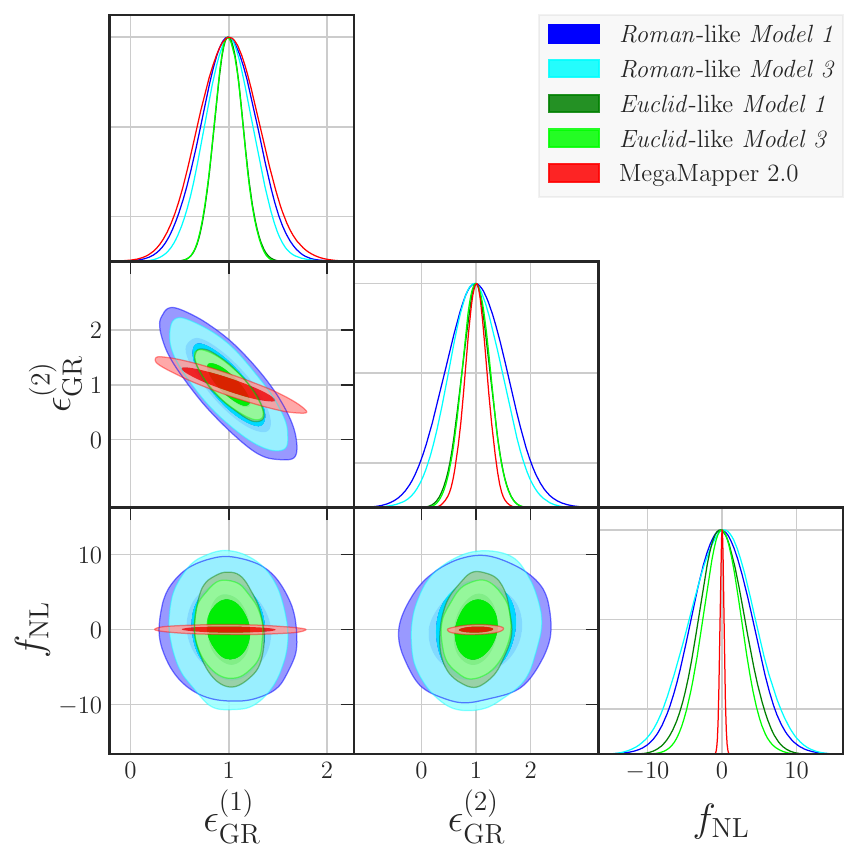}
\caption{Parameter constraints for each survey (we omit DESI BGS for clarity as the constraints are weak).}\label{fig:surv_contours}
\end{figure}

However, a significant improvement in precision on $f_{\mathrm{NL}}$ can be achieved at the high redshift range reached by MegaMapper LBG, giving a total $\sigma (f_{\mathrm{NL}})=0.31$. The total marginal errors are displayed in \cref{tab:errors} for fiducial $f_{\mathrm{NL}}=0$. The constraints on the parameters $\epsilon_{\rm GR}^{(1)} $, $\epsilon_{\rm GR}^{(2)} $ and $f_{\mathrm{NL}}$ can be seen in \cref{fig:surv_contours}. The improvement on constraints on $f_{\mathrm{NL}}$ from MegaMapper LBG is clearly shown in the bottom 3 panels, while the left panels show that we may be able to lift degeneracies between $\epsilon_{\rm GR}^{(1)}$ and $f_{\mathrm{NL}}$ with a combined MegaMapper/\textit{Euclid}-like survey. We note the degeneracy between $\epsilon_{\rm GR}^{(1)} $ and $\epsilon_{\rm GR}^{(2)} $, which may be due to our parametrisation in \cref{eq:B_GR}, where there are terms that are coupled to both amplitudes. This degeneracy may also be lifted with a MegaMapper/\textit{Euclid}-like combined analysis.

\begin{figure}[h!]
    \centering\includegraphics[width=.98\textwidth]{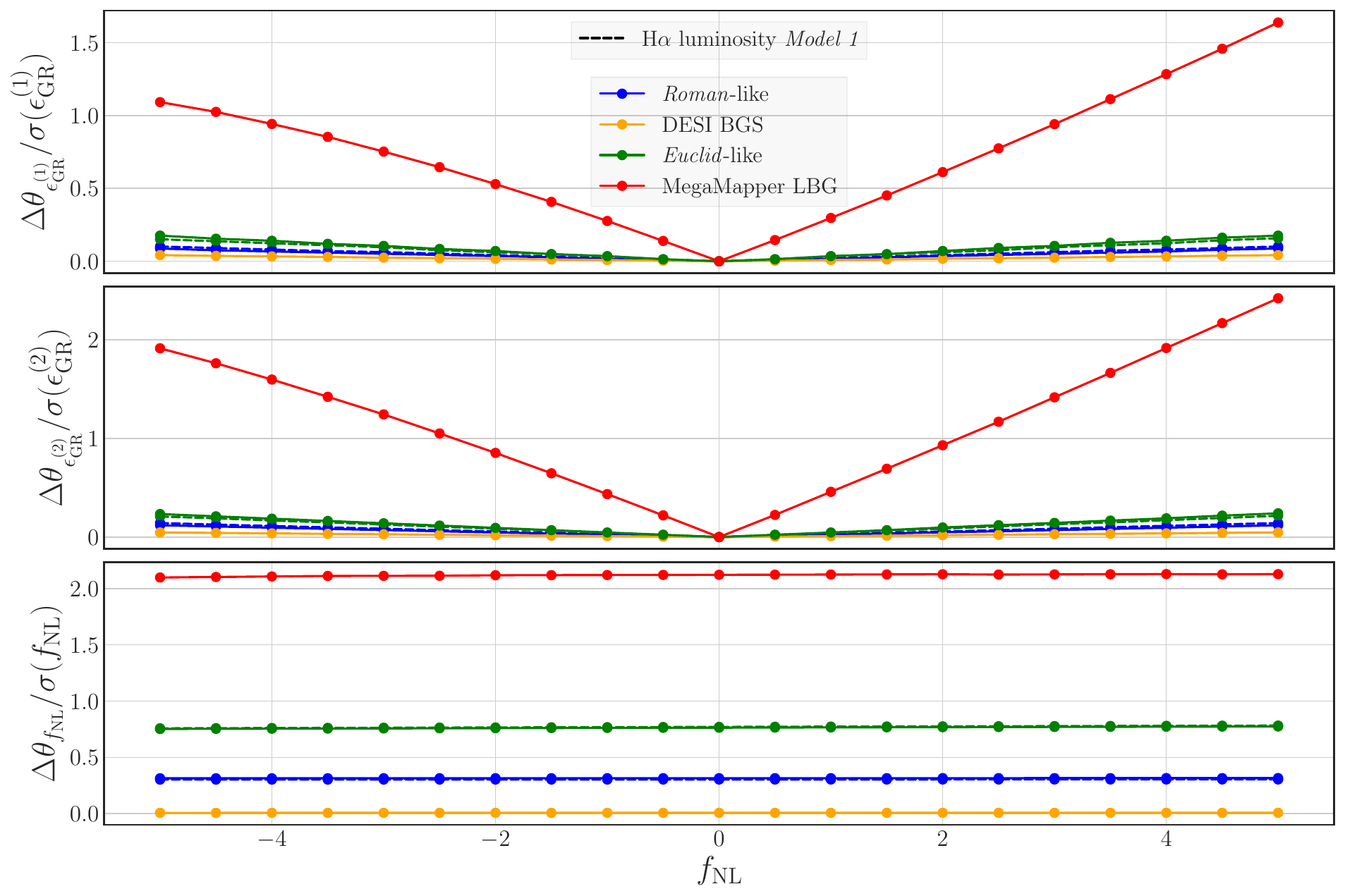}
    \caption{Total bias on parameter estimates for each survey, in units of marginal errors for $\epsilon_\mathrm{GR}^{(1)}$ (top), $\epsilon_\mathrm{GR}^{(2)}$ (centre) and $f_{\mathrm{NL}}$ (bottom) for $-5\leq f_{\mathrm{NL}}\leq 5$. }
    \label{fig:param bias}
\end{figure}

Implementing the parameter bias analysis of \cref{sec:param bias} for $-5<f_{\mathrm{NL}}<5$ yields \cref{fig:param bias}. We show the biases for the total information matrix, i.e.\ the sum over all redshift bins. 

 The top plot shows the bias that the assumed value of $f_{\mathrm{NL}}$, compared to the true value $f_{\mathrm{NL}}=0$, has on the measurements of the first-order GR amplitude $\epsilon_{\mathrm{GR}}^{(1)}$, in terms of the total marginal error. The middle plot shows the same for the second-order GR amplitude $\epsilon_{\mathrm{GR}}^{(2)}$, i.e.\ \cref{eq:grbias}. The bottom plot shows the bias on $f_{\mathrm{NL}}$ if GR corrections to the bispectrum are ignored, i.e.\ \cref{eq:fnlbias}. We can see that the impact on the measurement of both GR amplitudes $\epsilon_{\rm GR}^{(i)}$, where $i=1,2$, when $f_{\mathrm{NL}}$ is wrongly assumed, is $\lesssim 0.1 \sigma(\epsilon_{\rm GR})^{(i)}$, i.e.\ negligible for all surveys, apart from MegaMapper LBG, where the shifts are $\lesssim 1.5 \sigma(\epsilon_{\rm GR}^{(1)})$ and $\lesssim 2.1 \sigma(\epsilon_{\rm GR}^{(2)})$. Since these larger shift values are at the extremities of the $f_{\mathrm{NL}}$ range, this is not too concerning, since up to $f_{\mathrm{NL}}=\pm 2$, the shifts are $\lesssim 0.6 \sigma(\epsilon_{\rm GR}^{(1)})$ and $\lesssim 0.9 \sigma(\epsilon_{\rm GR}^{(2)})$. Conversely, from the bottom plot, we can see that precise and accurate measurements of $f_{\mathrm{NL}}$ rely on the inclusion of GR corrections in the analysis with surveys that probe the larger scales at higher redshift -- there is up to $2 \sigma(f_{\mathrm{NL}})$ discrepancy in measurement from the most promising survey results, when  GR effects are not included.

\begin{figure}[h!]
\centering
\includegraphics[width=.9\textwidth]{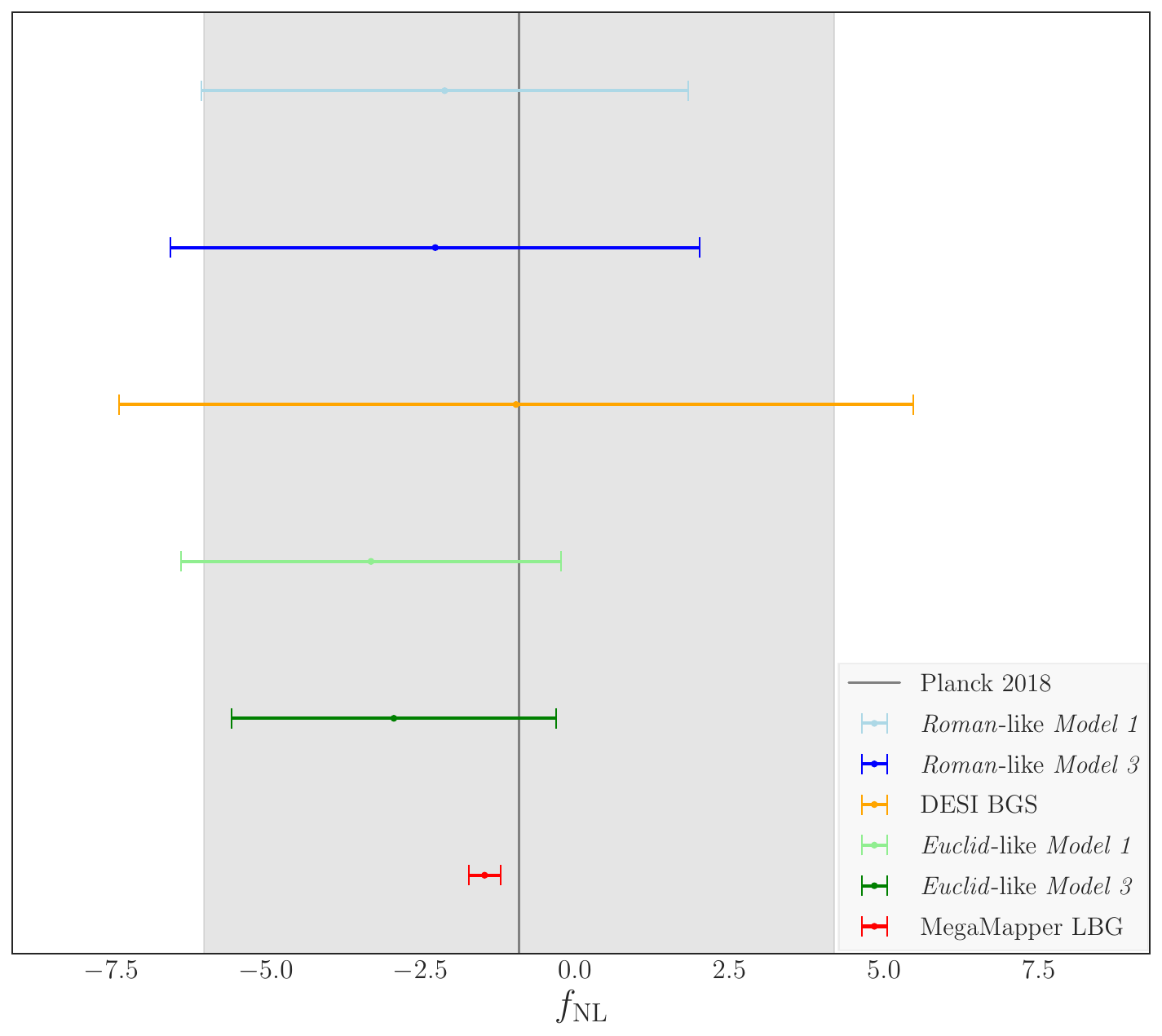}
\caption{Shift on the value of $f_{\mathrm{NL}}$ induced by neglecting GR corrections with error bars in the Newtonian regime. Here  the true value of $f_{\mathrm{NL}}$ is taken as that measured by Planck 2018 \cite{2020A&A...641A...9P}, $f_{\mathrm{NL}}=-0.9 \pm 5.1$, indicated by the grey vertical line and shaded area.}
\label{fig:fnl bias comp}
\end{figure}

This bias for each survey can be seen in \cref{fig:fnl bias comp}. Here we take the true value of $f_{\mathrm{NL}}=-0.9 \pm 5.1$ as measured by Planck 2018 \cite{2020A&A...641A...9P}, shown by the grey vertical line and shaded region representing the error. The shifted (or biased) values of $f_{\mathrm{NL}}$ for each survey are plotted along with the error bars given by the Fisher matrix of the form \cref{eq:marginal error} in the Newtonian regime, i.e.\ inserting \cref{eq:N bispectrum} for $B_{\rm gN}$ into \cref{eq:Fisher}, neglecting all GR and local PNG corrections. This plot demonstrates the importance of including GR corrections in constraining local PNG. The value of $f_{\mathrm{NL}}$ for a \textit{Euclid}-like survey is shifted to the edge of the margin of error given by Planck 2018, with error bars extending beyond it. From a quick glance we can see that the surveys with larger error bars have less $f_{\mathrm{NL}}$ bias induced in the Newtonian regime. However, the error bars are significantly larger, and extend outside the margin of error for Planck 2018. Once again, we can see how MegaMapper LBG does not follow this reasoning, due to its high signal detection of GR contributions and high accuracy on $f_{\mathrm{NL}}$. 

\subsection{Including luminosity derivatives} \label{sec:lum_derivs}
\begin{figure}[h!]
    \centering
    \includegraphics[width=.49\textwidth]{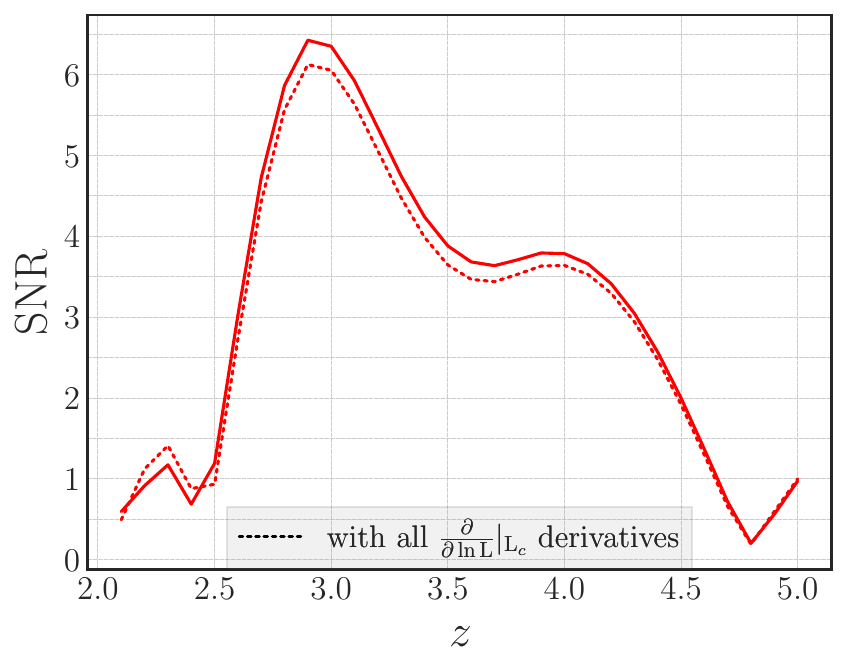}
    \includegraphics[width=.49\textwidth]{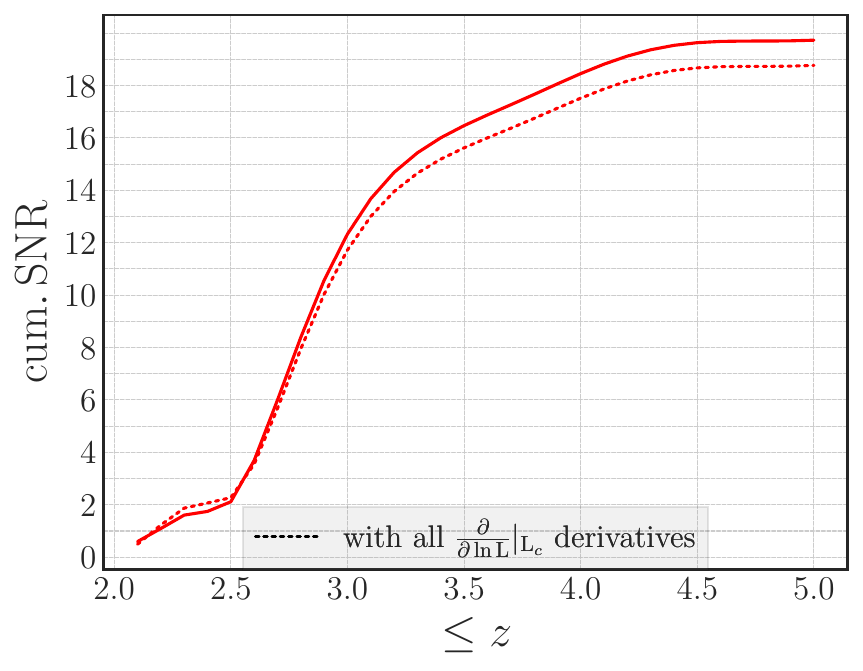}
    \caption{Comparison of total GR SNR for MegaMapper without luminosity derivatives (solid line, as shown in \cref{fig:surv_snrGR}) and with luminosity derivatives (dashed).}
    \label{fig:MMSNR}
\end{figure}

As mentioned in \cref{sec:GB}, the results above do not include some derivative terms in the $\beta$ and $\Upsilon$ functions of \cref{app:betas,app:upsilons}. Specifically, $\partial b_{10}/\partial\ln L$ and $\partial b_{01}/\partial\ln L$ (derivatives evaluated at luminosity cut $L_{\rm c}$). These are first-order ${\cal H}/k$ terms and were omitted as they require a clustering bias model which accounts for luminosity dependence. The clustering bias models we have for our chosen surveys do not satisfy this requirement, except in the case of MegaMapper, which we use to exemplify the incompleteness of the analysis without such terms. We  see from \cref{fig:MMSNR} that including these luminosity derivative terms in this case has a non-negligible effect on the overall signal of the GR contributions -- which suppresses the total GR correction.

\Cref{fig:MMbias} in \cref{app:MMB} shows that including these terms also has an effect on the bias on $f_{\mathrm{NL}}$. Consequently, it is crucial to include the luminosity derivative terms for a complete analysis. This highlights the need for the investigation of the clustering bias dependence on luminosity for different galaxy targets.

\section{Conclusion}
In this paper, we have investigated the detectability of relativistic corrections to the galaxy bispectrum in the presence of local PNG using analytically derived parameters of current and future galaxy surveys. The main conclusion that can be drawn from our analysis is that the higher the SNR of the GR signal, the more important it is to account for GR corrections when constraining $f_{\rm NL}$. This is shown most convincingly through \cref{fig:fnl bias comp}. The least promising survey from our analysis, with the lowest precision for all three parameters, is the DESI BGS, which appears to be ineffective at detecting any GR corrections to the galaxy bispectrum within a reasonable margin of error. While this means that there is little bias on $f_{\rm NL}$ when neglecting GR effects, it also means a huge uncertainty in the measurement of $f_{\rm NL}$ compared to samples at greater depth -- see the bottom left panel of  \cref{fig:comperrors}. 

Comparing the results for the \textit{Roman}- and \textit{Euclid}-like samples demonstrates that survey depth is more significant than sky coverage in the analysis of the interplay between relativistic effects and PNG. However, sky coverage does increase the signal of GR contributions and precision of the detection of all parameters. These two surveys are the fairest comparison of this since they target the same sample type and thus use the same luminosity functions. Therefore the only difference between them is the survey volume, or specifically survey sky area, since they probe similar redshift ranges. From \cref{fig:param bias,fig:fnl bias comp}, we see the bias on $f_{\rm NL}$ induced by the Newtonian approximation is far more significant than for DESI BGS, while the error bars are much smaller. Interestingly, \textit{Euclid}-like \textit{model 1} shows a larger shift on $f_{\rm NL}$ than \textit{Euclid}-like \textit{model 3}, as well as larger error bars. The smaller scope of a \textit{Roman}-like survey provides a lower SNR for the GR corrections which in turn gives a lower precision in the detection of their amplitudes (see \cref{fig:comperrors}). 

We find that MegaMapper LBG has a total GR SNR on a par with \textit{Euclid}-like \textit{model 3} (\cref{fig:surv_snrGR}). Although this is somewhat disappointing, given that MegaMapper LBG probes the scales where GR effects are most expected to mimic those of local PNG -- and a much larger range than the other surveys, it still indicates promising results. Of particular interest is the improvement on the detection of purely second-order GR contributions, seen in \cref{fig:surv_snrGR2}. Where MegaMapper LBG survey really outshines the other surveys, according to our study, is in measuring $f_{\mathrm{NL}}$ to high precision, as can be seen by the bottom two plots of \cref{fig:comperrors}. We double-checked this by computing the SNR for PNG corrections to the bispectrum, i.e.\ $B_{\rm NG}=B_{\rm g}- (B_{\rm gN}+ B_{\rm gGR})$ for $f_{\mathrm{NL}}=-0.9$, as in Planck 2018.  We find that the SNR for the PNG corrections to the MegaMapper LBG sample is about 4 times higher than that of \textit{Euclid}-like \textit{Model 3}.

This trend is further shown in computing the bias on $f_{\mathrm{NL}}$, as shown in \cref{fig:param bias}. For  MegaMapper LBG there is a $2\sigma$ shift on $f_{\rm NL}$ and a shift of less than $1\sigma$ for all other surveys. The implications of this are displayed clearly in \cref{fig:fnl bias comp}, where, despite the shift on \textit{Euclid}-like measurements of $f_{\rm NL}$ being very large, the true value lies within the Newtonian error bars. MegaMapper LBG shows a comparatively small shift on $f_{\rm NL}$, however the shift is almost double that of the precision of the survey. Thus, in a Newtonian analysis, the true value of $f_{\rm NL}$ would not lie within the margin of error of the observed value for the survey. This implies that for MegaMapper and future surveys with similar capability, any attempt to measure and constrain $f_{\rm NL}$ without accounting for GR corrections, would be futile.

A particularly promising finding of our analysis is the combined constraining power of MegaMapper LBG and \textit{Euclid}-like H$\alpha$ samples. The bottom left panel of \cref{fig:surv_contours} displays how MegaMapper's sensitivity to $f_{\rm NL}$ can be paired with Euclid's slight sensitivity to $\epsilon_{GR}^{(1)}$ to break degeneracies between the two parameters. Similarly, the surveys may complement each other to break degeneracies between $\epsilon_{GR}^{(1)}$ and $\epsilon_{GR}^{(2)}$, as can be seen in the middle left panel. It would not be unreasonable to expect a larger cumulative GR signal from a MegaMapper LBG sample since it probes a large range within the high redshift regime, capturing the ultra-large scales beyond the equality scale, on which relativistic effects are most important and primordial non-Gaussianity can be probed most accurately.

There are a few reasons why MegaMapper may be less successful in our analysis than we anticipated. Firstly, we believe that the fluctuation of the evolution bias $\mathcal E$ from positive to negative values (see \cref{app:sp,fig:E}), as derived from the luminosity model described in \cref{app:MMB}, may cause a significant suppression. The imprint of the evolution bias is evident for all the surveys; for example, the H$\alpha$ \textit{Model 1} discontinuity appears clearly in both \textit{Roman} and \textit{Euclid}-like \textit{Model 1} SNR analyses. Similarly for MegaMapper LBG, the peaks and troughs in the evolution bias are almost completely mirrored in the SNR. Additionally, the redshift bins in which $\mathcal E$ is negative, $2.6 \lesssim z \lesssim 4.6 $, produce the highest signal -- and where it is positive, the signal is significantly suppressed.

Secondly, our exclusion of some relativistic effects may  undermine the true potential of MegaMapper's capabilities. The dominant effect in our model is the Doppler correction \cite{Maartens:2019yhx}. This is a reflection of the peculiar velocities of galaxies which are significantly larger at lower redshift due to the local clustering of galaxies. As we move to higher redshift ranges, the Doppler correction becomes weaker and lensing magnification, an integrated GR effect which we have not included in our study, becomes more important. If we were to include this, as well as integrated Sachs-Wolf  and time delay effects which accumulate along the line-of-sight and thus become stronger at higher redshift -- we may obtain a significantly higher signal and thus achieve a higher precision
on the GR amplitudes that match the constraining power that has been shown for $f_{\mathrm{NL}}$.

It may also be worth exploring other parametrisations when introducing the arbitrary amplitudes to the bispectrum, as our method involves cross-amplitude terms which may exacerbate degeneracies between them. Further improvements may be achieved through developments of determining the comoving galaxy number density, i.e.\ improved modelling of the luminosity function or using halo occupation distributions from simulations. This in turn may reduce the suppression of the evolution bias and allow for higher SNR of the GR corrections to the galaxy bispectrum.

This work is the first to forecast constraints on relativistic amplitudes and $f_{\rm NL}$ using an analytical model of the observed galaxy bispectrum with local projection effects in the presence of local PNG. We aim to highlight the information and constraining power on $f_{\rm NL}$ which can be obtained uniquely through bispectrum measurements within the coming data releases from galaxy surveys, if relativistic effects are carefully modelled. A multi-tracer and a joint $P+B$ approach is left for future work.


\acknowledgments
SR and SC acknowledge support from the Italian Ministry of University and Research (\textsc{mur}), PRIN 2022 `EXSKALIBUR – Euclid-Cross-SKA: Likelihood Inference Building for Universe's Research', Grant No.\ 20222BBYB9, CUP D53D2300252 0006, from the Italian Ministry of Foreign Affairs and International
Cooperation (\textsc{maeci}), Grant No.\ ZA23GR03, and from the European Union -- Next Generation EU. RM is supported by the South African Radio Astronomy Observatory and the National Research Foundation (Grant No. 75415).

\newpage
\appendix
\newpage
\section{Fourier kernels and biases} 
\subsection{Fourier kernels}\label{app:kernels}
\begin{align}
F_2\left(\boldsymbol{k}_1, \boldsymbol{k}_2\right)&=1+\frac{F}{D^2}+\left(\frac{k_1}{k_2}+\frac{k_2}{k_1}\right)\,\hat{\boldsymbol{k}}_1 \cdot \hat{\boldsymbol{k}}_2+\left[1-\frac{F}{D^2}\right]\,\left(\hat{\boldsymbol{k}}_1 \cdot \hat{\boldsymbol{k}}_2\right)^2,\\
G_2\left(\boldsymbol{k}_1, \boldsymbol{k}_2\right)&=\frac{F^{\prime}}{D\,D^{\prime}}+\left(\frac{k_1}{k_2}+\frac{k_2}{k_1}\right)\,\hat{\boldsymbol{k}}_1 \cdot \hat{\boldsymbol{k}}_2+\left(2-\frac{F^{\prime}}{D\,D^{\prime}}\right)\,\left(\hat{\boldsymbol{k}}_1 \cdot \hat{\boldsymbol{k}}_2\right)^2,\\
S_2\left(\boldsymbol{k}_1, \boldsymbol{k}_2\right)&=\left(\hat{\boldsymbol{k}}_1 \cdot \hat{\boldsymbol{k}}_2\right)^2-\frac{1}{3}\;, \\
N^{(2)}\left(\boldsymbol{k}_1, \boldsymbol{k}_2\right)&=\frac{1}{2}\,\left(\frac{k_1}{k_2\,\mathcal{M}_1}+\frac{k_2}{k_1\,\mathcal{M}_2}\right)\,\hat{\boldsymbol{k}}_1 \cdot \hat{\boldsymbol{k}}_2 \;,
\end{align}
where we use the Einstein-de Sitter relations $F/D^2=3/7$ and $F^{\prime}/(D\,D^{\prime})=6/7$ as a reasonable approximation for $\Lambda\mathrm{CDM}$.

\subsection{Gaussian clustering biases}\label{app:g biases}

\begin{align}
    \textbf{\textit{Euclid/Roman}:}\;
    b_{10}&= 0.9 + 0.4\,z\\
    b_{20}&= -0.741 - 0.125\,z + 0.123\, z^2 + 0.00637\,z^3\\
    \nonumber \\
    \mathbf{DESI \;BGS:}\;    b_{10}&= 1.34/D(z)\\  b_{20}&=0.30-0.79\,b_{10}+0.20\,b_{10}^2+0.12/b_{10}\\
    \nonumber \\\mathbf{MegaMapper \; LBG:}\label{eq:bMM}\;b_{10}&=0.49\,(1+z) + 0.11\,(1+z)^2 \\ b_{20}&=0.30-0.79\, b_{10}+0.20 \,b_{10}^2+0.12/b_{10}
\end{align}

\subsection{Non-Gaussian biases}\label{nG biases}
\begin{align}
b_{01}&=2\,f_{\mathrm{NL}}\,\delta_{\text {crit }}\,\left(b_{10}-1\right) \\
b_{11} & =4\,f_{\mathrm{NL}}\,\left[\delta_{\text {crit }}\,b_{20}+\left(\frac{13}{21}\,\delta_{\text {crit }}-1\right)\,\left(b_{10}-1\right)+1\right], \\
b_N & =4\,f_{\mathrm{NL}}\,\left[\delta_{\text {crit }}\,\left(1-b_{10}\right)+1\right], \\
b_{02} & =4\,f_{\mathrm{NL}}^2\,\delta_{\text {crit }}\,\left[\delta_{\text {crit }} b_{20}-2\,\left(\frac{4}{21}\,\delta_{\text {crit }}+1\right)\,\left(b_{10}-1\right)\right] .
\end{align}
\section{$\beta_I$ functions}\label{app:betas}
We list the $\beta_I$ functions of \cref{eq:KGR2}. Note that here and in \autoref{app:upsilons}, $L$ denotes the limiting luminosity. Also note that all partial derivatives with respect to $L$ have been neglected in the analysis, with the exception of the discussion in \cref{sec:lum_derivs}, to which we refer for more details.
\begin{align}
 \frac{\beta_1}{\mathcal{H}^4}&=\frac{9}{4} \Omega_{\rm m}^2\left[6-2\,f\left(2\mathcal{E}-4\,\mathcal{Q}-\frac{4\,(1-\mathcal{Q})}{\chi\,\mathcal{H}}-\frac{2 \mathcal{H}^{\prime}}{\mathcal{H}^2}\right)-\frac{2\,f^{\prime}}{\mathcal{H}}+\mathcal{E}^2+5\,\mathcal{E}-8\,\mathcal{E}\,\mathcal{Q}+4\,\mathcal{Q}+16\,\mathcal{Q}^2\right. \nonumber \\
& \quad -16 \frac{\partial\mathcal{Q}}{\partial \ln L}-8 \frac{\mathcal{Q}^{\prime}}{\mathcal{H}}+\frac{\mathcal{E}^{\prime}}{\mathcal{H}}+\frac{2}{\chi^2 \mathcal{H}^2}\left(1-\mathcal{Q}+2\,\mathcal{Q}^2-2 \frac{\partial\mathcal{Q}}{\partial \ln L}\right) \nonumber \\
& \quad -\frac{2}{\chi\,\mathcal{H}}\left(3+2\,\mathcal{E}-2\,\mathcal{E}\,\mathcal{Q}-3\,\mathcal{Q}+8\,\mathcal{Q}^2-\frac{3 \mathcal{H}^{\prime}}{\mathcal{H}^2}\,(1-\mathcal{Q})-8 \frac{\partial\mathcal{Q}}{\partial \ln L}-2 \frac{\mathcal{Q}^{\prime}}{\mathcal{H}}\right)\nonumber \\
& \quad \left.+\frac{\mathcal{H}^{\prime}}{\mathcal{H}^2}\left(-7-2\,\mathcal{E}+8\,\mathcal{Q}+\frac{3 \mathcal{H}^{\prime}}{\mathcal{H}^2}\right)-\frac{\mathcal{H}^{\prime \prime}}{\mathcal{H}^3}\right]\nonumber \\
& \quad +\frac{3}{2} \Omega_{\rm m} f\left[5-2\,f\left(4-\mathcal{E}\right)+\frac{2\,f^{\prime}}{\mathcal{H}}+2\,\mathcal{E}\left(5+\frac{2\,(1-\mathcal{Q})}{\chi{\mathcal{H}}}\right)-\frac{2\,\mathcal{E}^{\prime}}{\mathcal{H}}-2\,\mathcal{E}^2+8\,\mathcal{E}\,\mathcal{Q}-28\,\mathcal{Q}\right. \nonumber\\
& \quad \left.-\frac{14\,(1-\mathcal{Q})}{\chi\,\mathcal{H}}-\frac{3 \mathcal{H}^{\prime}}{\mathcal{H}^2}+4\left(2-\frac{1}{\chi\,\mathcal{H}}\right) \frac{\mathcal{Q}^{\prime}}{\mathcal{H}}\right] \nonumber\\
& \quad +\frac{3}{2} \Omega_{\rm m} f^2\left[-2+2\,f-\mathcal{E}+4\,\mathcal{Q}+\frac{2\,(1-\mathcal{Q})}{\chi\,\mathcal{H}}+\frac{3 \mathcal{H}^{\prime}}{\mathcal{H}^2}\right]\nonumber \\
& \quad +f^2\left[12-7\,\mathcal{E}+\mathcal{E}^2+\frac{\mathcal{E}^{\prime}}{\mathcal{H}}+\left(\mathcal{E}-3\right) \frac{\mathcal{H}^{\prime}}{\mathcal{H}^2}\right]-\frac{3}{2} \Omega_{\rm m} \frac{f^{\prime}}{\mathcal{H}} \\
\frac{\beta_2}{\mathcal{H}^4}&=\frac{9}{2} \Omega_{\rm m}^2\left[-1+\mathcal{E}-2\,\mathcal{Q}-\frac{2\,(1-\mathcal{Q})}{\chi\,\mathcal{H}}-\frac{\mathcal{H}^{\prime}}{\mathcal{H}^2}\right]+3 \Omega_{\rm m} f\left[-1+2\,f-\mathcal{E}+4\,\mathcal{Q}+\frac{2\,(1-\mathcal{Q})}{\chi\,\mathcal{H}}+\frac{3 \mathcal{H}^{\prime}}{\mathcal{H}^2}\right] \nonumber\\
& \quad +3 \Omega_{\rm m} f^2\left[-1+\mathcal{E}-2\,\mathcal{Q}-\frac{2\,(1-\mathcal{Q})}{\chi\,\mathcal{H}}-\frac{\mathcal{H}^{\prime}}{\mathcal{H}^2}\right]+3 \Omega_{\rm m} \frac{f^{\prime}}{\mathcal{H}} \\
\frac{\beta_3}{\mathcal{H}^3}&=\frac{9}{4} \Omega_{\rm m}^2(f-2+2\,\mathcal{Q})+\frac{3}{2} \Omega_{\rm m} f\left[-2-f\left(-3+f+2\,\mathcal{E}-3\,\mathcal{Q}-\frac{4\,(1-\mathcal{Q})}{\chi\,\mathcal{H}}-\frac{2 \mathcal{H}^{\prime}}{\mathcal{H}^2}\right)-\frac{f^{\prime}}{\mathcal{H}}\right.\nonumber \\
& \quad +3\,\mathcal{E}+\mathcal{E}^2-6\,\mathcal{E}\,\mathcal{Q}+4\,\mathcal{Q}+8\,\mathcal{Q}^2-8 \frac{\partial\mathcal{Q}}{\partial \ln L}-6 \frac{\mathcal{Q}^{\prime}}{\mathcal{H}}+\frac{\mathcal{E}^{\prime}}{\mathcal{H}} \nonumber\\
& \quad +\frac{2}{\chi^2 \mathcal{H}^2}\left(1-\mathcal{Q}+2\,\mathcal{Q}^2-2 \frac{\partial\mathcal{Q}}{\partial \ln L}\right)+\frac{2}{\chi\,\mathcal{H}}\left(-1-2\,\mathcal{E}+2\,\mathcal{E}\,\mathcal{Q}+\mathcal{Q}-6\,\mathcal{Q}^2\right. \nonumber\\
&\quad  \left.\left.+\frac{3 \mathcal{H}^{\prime}}{\mathcal{H}^2}\,(1-\mathcal{Q})+6 \frac{\partial\mathcal{Q}}{\partial \ln L}+2 \frac{\mathcal{Q}^{\prime}}{\mathcal{H}}\right)-\frac{\mathcal{H}^{\prime}}{\mathcal{H}^2}\left(3+2\,\mathcal{E}-6\,\mathcal{Q}-\frac{3 \mathcal{H}^{\prime}}{\mathcal{H}^2}\right)-\frac{\mathcal{H}^{\prime \prime}}{\mathcal{H}^3}\right] \nonumber\\
& \quad +f^2\left[-3+2\,\mathcal{E}\left(2+\frac{\,(1-\mathcal{Q})}{\chi\,\mathcal{H}}\right)-\mathcal{E}^2+2\,\mathcal{E}\,\mathcal{Q}-6\,\mathcal{Q}-\frac{\mathcal{E}^{\prime}}{\mathcal{H}}-\frac{6\,(1-\mathcal{Q})}{\chi\,\mathcal{H}}+2\left(1-\frac{1}{\chi\,\mathcal{H}}\right) \frac{\mathcal{Q}^{\prime}}{\mathcal{H}}\right] \\
\frac{\beta_4}{\mathcal{H}^3}&=\frac{9}{2} \Omega_{\rm m} f\left[-\mathcal{E}+2\,\mathcal{Q}+\frac{2\,(1-\mathcal{Q})}{\chi\,\mathcal{H}}+\frac{\mathcal{H}^{\prime}}{\mathcal{H}^2}\right] \\
\frac{\beta_5}{\mathcal{H}^3}&=3 \Omega_{\rm m} f\left[\mathcal{E}-2\,\mathcal{Q}-\frac{2\,(1-\mathcal{Q})}{\chi\,\mathcal{H}}-\frac{\mathcal{H}^{\prime}}{\mathcal{H}^2}\right] \\
\frac{\beta_6}{\mathcal{H}^2}&=\frac{3}{2} \Omega_{\rm m}\left[2-2\,f+\mathcal{E}-4\,\mathcal{Q}-\frac{2\,(1-\mathcal{Q})}{\chi\,\mathcal{H}}-\frac{\mathcal{H}^{\prime}}{\mathcal{H}^2}\right] \\
\frac{\beta_7}{\mathcal{H}^2}&=f\left(3-\mathcal{E}\right) \\
\nonumber\\
\frac{\beta_8}{\mathcal{H}^2}&=3 \Omega_{\rm m} f(2-f-2\,\mathcal{Q})+f^2\left[4+\mathcal{E}-\mathcal{E}^2+4\,\mathcal{E}\,\mathcal{Q}-6\,\mathcal{Q}-4\,\mathcal{Q}^2+4 \frac{\partial\mathcal{Q}}{\partial \ln L}+4 \frac{\mathcal{Q}^{\prime}}{\mathcal{H}}-\frac{\mathcal{E}^{\prime}}{\mathcal{H}}\right. \nonumber \\
& \quad -\frac{2}{\chi^2 \mathcal{H}^2}\left(1-\mathcal{Q}+2\,\mathcal{Q}^2-2 \frac{\partial\mathcal{Q}}{\partial \ln L}\right)-\frac{2}{\chi\,\mathcal{H}}\left(3-2\,\mathcal{E}+2\,\mathcal{E}\,\mathcal{Q}-\mathcal{Q}-4\,\mathcal{Q}^2+\frac{3 \mathcal{H}^{\prime}}{\mathcal{H}^2}\,(1-\mathcal{Q})\right.\nonumber \\
& \quad \left.\left.+4 \frac{\partial\mathcal{Q}}{\partial \ln L}+2 \frac{\mathcal{Q}^{\prime}}{\mathcal{H}}\right)-\frac{\mathcal{H}^{\prime}}{\mathcal{H}^2}\left(3-2\,\mathcal{E}+4\,\mathcal{Q}+\frac{3 \mathcal{H}^{\prime}}{\mathcal{H}^2}\right)+\frac{\mathcal{H}^{\prime \prime}}{\mathcal{H}^3}\right] \\
\frac{\beta_9}{\mathcal{H}^2}&=-\frac{9}{2} \Omega_{\rm m} f \\
\frac{\beta_{10}}{\mathcal{H}^2}&=3 \Omega_{\rm m} f \\
\frac{\beta_{11}}{\mathcal{H}^2}&=\frac{3}{2} \Omega_{\rm m}\left(1+\frac{2\,f}{3 \Omega_{\rm m}}\right)+3 \Omega_{\rm m} f-f^2\left[-1+\mathcal{E}-2\,\mathcal{Q}-\frac{2\,(1-\mathcal{Q})}{\chi\,\mathcal{H}}-\frac{\mathcal{H}^{\prime}}{\mathcal{H}^2}\right] \\
\frac{\beta_{12}}{\mathcal{H}^2}&=-3 \Omega_{\rm m}\left(1+\frac{2\,f}{3 \Omega_{\rm m}}\right)-f\left[b_{10}\left(f-3+\mathcal{E}\right)+\frac{b_{10}^{\prime}}{\mathcal{H}}\right] \nonumber \\
& \quad +\frac{3}{2} \Omega_{\rm m}\left[b_{10}\left(2+\mathcal{E}-4\,\mathcal{Q}-\frac{2\,(1-\mathcal{Q})}{\chi\,\mathcal{H}}-\frac{\mathcal{H}^{\prime}}{\mathcal{H}^2}\right)+\frac{b_{10}^{\prime}}{\mathcal{H}}+2\left(2-\frac{1}{\chi\,\mathcal{H}}\right) \frac{\partial b_{10}}{\partial \ln L}\right] \\
\frac{\beta_{13}}{\mathcal{H}^2}&=\frac{9}{4} \Omega_{\rm m}^2+\frac{3}{2} \Omega_{\rm m} f\left[1-2\,f+2\,\mathcal{E}-6\,\mathcal{Q}-\frac{4\,(1-\mathcal{Q})}{\chi\,\mathcal{H}}-\frac{3 \mathcal{H}^{\prime}}{\mathcal{H}^2}\right]+f^2\left(3-\mathcal{E}\right) \\
\frac{\beta_{14}}{\mathcal{H}}&=-\frac{3}{2} \Omega_{\rm m} b_{10} \\
\frac{\beta_{15}}{\mathcal{H}}&=2\,f^2 \\
\frac{\beta_{16}}{\mathcal{H}}&=f\left[b_{10}\left(f+\mathcal{E}-2\,\mathcal{Q}-\frac{2\,(1-\mathcal{Q})}{\chi\,\mathcal{H}}-\frac{\mathcal{H}^{\prime}}{\mathcal{H}^2}\right)+\frac{b_{10}^{\prime}}{\mathcal{H}}+2\left(1-\frac{1}{\chi\,\mathcal{H}}\right) \frac{\partial b_{10}}{\partial \ln L}\right] \\
\frac{\beta_{17}}{\mathcal{H}}&=-\frac{3}{2} \Omega_{\rm m} f \\
\frac{\beta_{18}}{\mathcal{H}}&=\frac{3}{2} \Omega_{\rm m} f-f^2\left[3-2\,\mathcal{E}+4\,\mathcal{Q}+\frac{4\,(1-\mathcal{Q})}{\chi\,\mathcal{H}}+\frac{3 \mathcal{H}^{\prime}}{\mathcal{H}^2}\right] \\
\frac{\beta_{19}}{\mathcal{H}}&=f\,\left[\mathcal{E}-2\,Q-\frac{2\,(1-\mathcal{Q})}{\chi\,\mathcal{H}}-\frac{\mathcal{H}^{\prime}}{\mathcal{H}^2}\right] 
\end{align}
\section{$\Upsilon_I$ functions}\label{app:upsilons}
\begin{align}
\frac{1}{f_{\mathrm{NL}}} \frac{\Upsilon_1}{\mathcal{H}^2} &=  2\left(3-\mathcal{E}\right) f+3 \Omega_{\rm m}\left[1+\mathcal{E}-4\,\mathcal{Q}-\frac{2\,(1-\mathcal{Q})}{\chi\,\mathcal{H}}-\frac{\mathcal{H}^{\prime}}{\mathcal{H}^2}\right] \nonumber\\
& \quad +\frac{6 \Omega_{\rm m}}{\left(3 \Omega_{\rm m}+2\,f\right)}\left[\frac{f^{\prime}}{\mathcal{H}}+\left(1+2 \frac{\mathcal{H}^{\prime}}{\mathcal{H}^2}\right) f\right] \\
\frac{1}{f_{\mathrm{NL}}} \frac{\Upsilon_2}{\mathcal{H}}&= 2\,f\left[\mathcal{E}-2\,\mathcal{Q}-\frac{2\,(1-\mathcal{Q})}{\chi\,\mathcal{H}}-\frac{\mathcal{H}^{\prime}}{\mathcal{H}^2}\right] \\
\frac{1}{b_{01}} \frac{\Upsilon_3}{\mathcal{H}^2}&=  \frac{3}{2} \Omega_{\rm m}\left[2+\mathcal{E}-4\,\mathcal{Q}+\frac{2\,(1-\mathcal{Q})}{\chi\,\mathcal{H}}+\frac{\mathcal{H}^{\prime}}{\mathcal{H}^2}+2\left(2-\frac{1}{\chi\,\mathcal{H}}\right) \frac{\partial \ln b_{01}}{\partial \ln L}\right] \nonumber\\
& \quad +f\left[3-f-\mathcal{E}+\frac{1}{2} \frac{\partial \ln b_{01}}{\partial \ln a}\right] \\
\frac{1}{b_{01}} \frac{\Upsilon_4}{\mathcal{H}}&= -\frac{3}{2} \Omega_{\rm m} \\
\frac{1}{b_{01}} \frac{\Upsilon_5}{\mathcal{H}}&=  f\left[f+\mathcal{E}-2\,\mathcal{Q}-\frac{2\,(1-\mathcal{Q})}{\chi\,\mathcal{H}}-\frac{\mathcal{H}^{\prime}}{\mathcal{H}^2}+2\left(2-\frac{1}{\chi\,\mathcal{H}}\right) \frac{\partial \ln b_{01}}{\partial \ln L}\right]
\end{align}
\clearpage
\newpage
\section{MegaMapper LBG biases}\label{app:MMB}
In order to derive the comoving galaxy number density $n_{\mathrm{g}}$, magnification bias $\mathcal{Q}$ and evolution bias $\mathcal{E}$ for the MegaMapper LBG sample we need the luminosity function. We use the Schechter type UV luminosity function with absolute magnitude \cite{Sailer:2021yzm,Wilson:2019brt}
\begin{equation}\label{eq:MMlum}
   \Phi\left(M_{\mathrm{UV}}\right)\,\de M_{\mathrm{UV}}=\left(\frac{\ln 10}{2.5}\right)\,\phi^{\star}\,10^{-0.4\,(1+\alpha)\,\left(M_{\mathrm{UV}}-M_{\mathrm{UV}}^\star\right)}\,\exp\left[-10^{-0.4\,\left(M_{\mathrm{UV}}-M_{\mathrm{UV}}^\star\right)}\right]\,\de M_{\mathrm{UV}}\;,
\end{equation}
where
\begin{align}\label{eq:Muv}
    M_{\rm UV}&=m-5\,\log _{10}\left[\frac{D_{\rm L}(z)}{10\,\mathrm{pc}}\right] - K(z)\;\\
\label{eq:K-corr}
     K(z) &= - 2.5\,\log _{10}(1+z)- \big(m_{\rm UV}-m \big).
 \end{align}
\cref{eq:K-corr} is the $K$-correction which accounts for the redshifting effect on galaxy fluxes which are measured in fixed wavelength bands. $m_{\rm UV} - m$ is negligible for the MegaMapper LBG sample, so we choose to ignore it. $D_{\rm L}(z)$ is the luminosity distance and the best-fit parameters for equation \cref{eq:MMlum} are given by Table~3 of \citep{Wilson:2019brt}[reported also here in \cref{tab:MM lum params}]. This paper gives a full description of the properties of LBGs and their selection process. 

\begin{table}[h!]
    \centering
    \caption{Luminosity function best-fit parameters as given by Ref.\ \cite{Wilson:2019brt}, which gives references for deriving the best-fit values at each redshift.
    \label{tab:MM lum params}}
    \begin{tabular}{cccccc}
    \hline$z_{{\rm eff}}$ & $M_{\rm UV}^\star$ & $\phi^* /\left(10\,h^{-1}\,\mathrm{Mpc}\right)^{-3}$ & $\alpha$ & $m_{\rm UV}^\star$ \\
    \hline
    \hline $2.0$ & $-20.60$ & $9.70$ & $-1.60$ & $24.2$ \\
    \hline $3.0$ & $-20.86$ & $5.04$ & $-1.78$ & $24.7$ \\
    \hline $3.8$ & $-20.63$ & $9.25$ & $-1.57$ & $25.4$  \\
    \hline $4.9$ & $-20.96$ & $3.22$ & $-1.60$ & $25.5$  \\
    \hline $5.9$ & $-20.91$ & $1.64$ & $-1.87$ & $25.8$  \\
    \hline
    \end{tabular}
\end{table}

\begin{figure}[h!]
\centering
\begin{subfigure}[b]{.45\textwidth}
   \includegraphics[width=\textwidth]{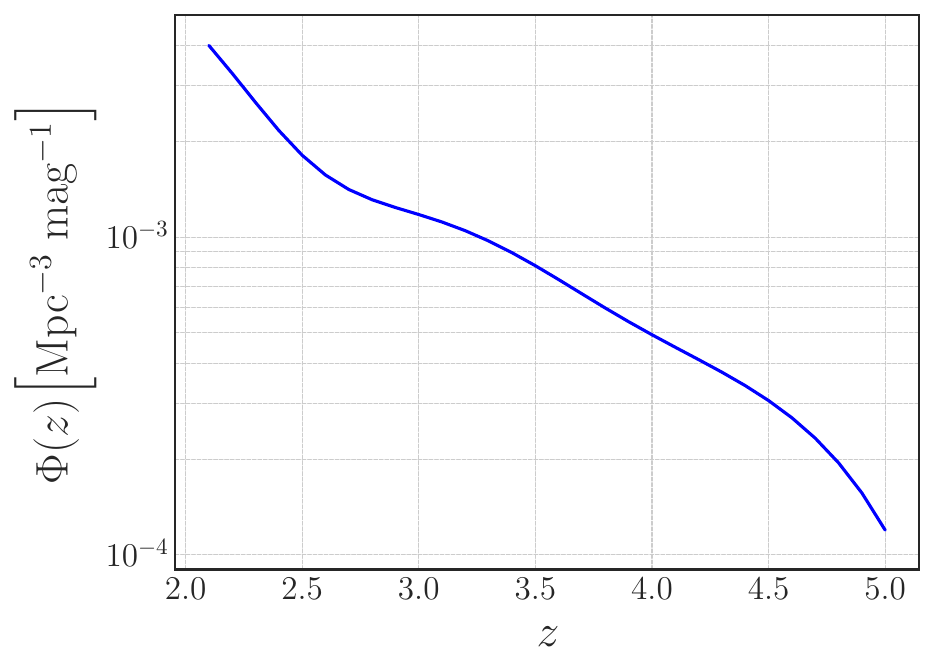}
   \caption{Luminosity function $\Phi$}
   \label{fig:lumfunc} 
\end{subfigure}
\hfill
\begin{subfigure}[b]{.45\textwidth}
    \includegraphics[width=\textwidth]{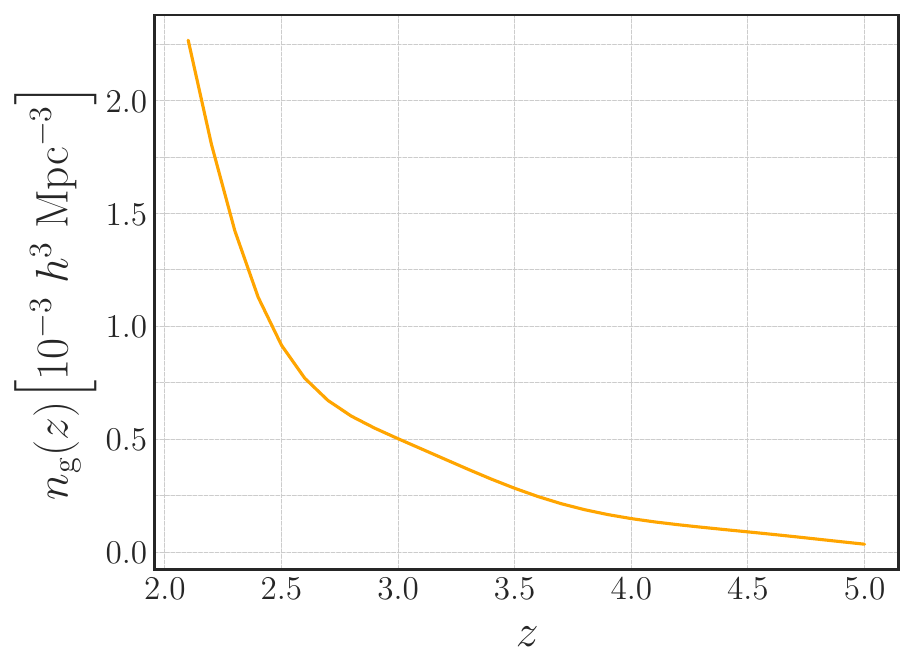}
    \caption{Comoving number density $n_{\rm g}$}
    \label{fig:ng}
\end{subfigure}
\qquad
\begin{subfigure}[b]{.45\textwidth}
    \includegraphics[width=\textwidth]{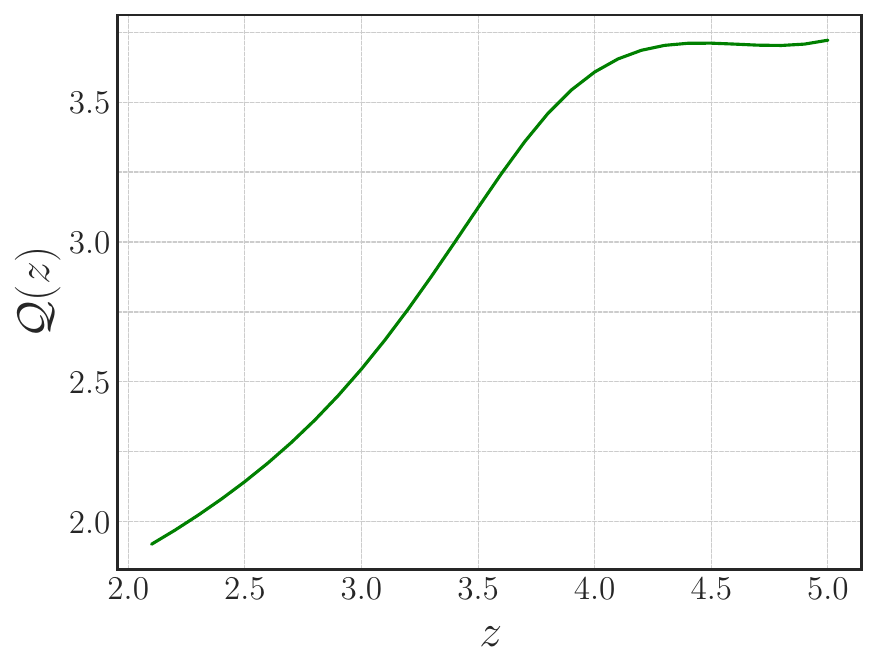}
    \caption{Magnification bias $\mathcal{Q}$}
    \label{fig:Q}  
\end{subfigure}
\hfill
\begin{subfigure}[b]{.45\textwidth}
    \includegraphics[width=\textwidth]{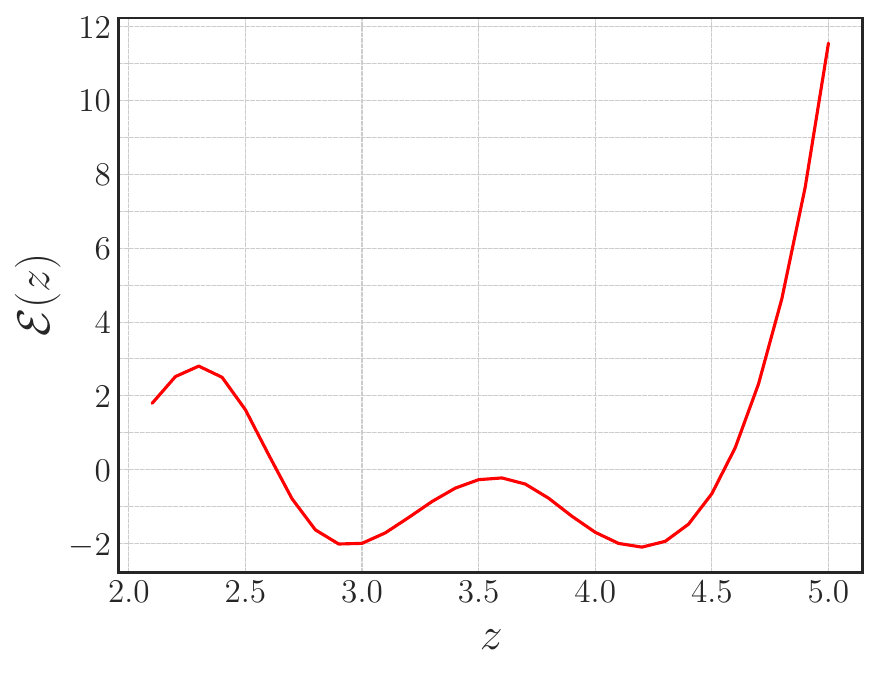}
    \caption{Evolution bias $\mathcal{E}$}
    \label{fig:E}
\end{subfigure}
\caption{MegaMapper LBG luminosity function, comoving redshift distribution, and magnification and evolution biases derived from Ref.\ \cite{Wilson:2019brt, Maartens:2021dqy}.}
\label{fig:MMparams}
\end{figure}

As the \cref{tab:MM lum params} values are sparse across the redshift range, we use a cubic spline function to smooth them. This gives the luminosity function  in \cref{fig:lumfunc}. In order to derive the comoving number density, evolution bias and magnification bias, we follow \cite{Maartens:2021dqy}, specifically the section on surveys with $K$-corrections applied to a DESI-like survey. The comoving number density is given by (see \cref{fig:ng}),
\begin{equation}
    n_{\mathrm{g}}\left(z, M_{\mathrm{c}}\right)=\int_{-\infty}^{M_{\mathrm{c}}(z)}\,\de M\,\Phi(z, M),
\end{equation}
where $M_{\mathrm{c}}$ is given by \cref{eq:Muv} for $m=m_{\mathrm{c}}=24.5$. The magnification bias is given by the relation (see \cref{fig:Q}),
\begin{equation}
   \mathcal{Q}\,\left(z, M_{\mathrm{c}}\right)=\frac{5}{2}\,\frac{\partial \log _{10} n_{\mathrm{g}}\,\left(z, M_{\mathrm{c}}\right)}{\partial M_{\mathrm{c}}}=\frac{5}{2\,\ln 10}\,\frac{\Phi\left(z, M_{\mathrm{c}}\right)}{n_{\mathrm{g}}\left(z, M_{\mathrm{c}}\right)},
\end{equation}
 and the evolution bias is calculated by (see \cref{fig:E})
\begin{equation}
   \mathcal{E}=-\frac{\de \ln n_{\mathrm{g}}}{\de \ln (1+z)}-2\,\left[1+\frac{1}{\chi\,{\cal H}}+\frac{2\,\ln 10}{5}\,\frac{\de K}{\de\ln(1+z)}\right]\,\mathcal{Q}\;.
\end{equation}
We continue to follow \cite{Wilson:2019brt} and \cite{Sailer:2021yzm} by using the same linear galaxy bias model:
\begin{equation}
    b_{10}=A(m)\,(1+z) + B(m_c)\,(1+z)^2\; ,
\end{equation}
with 
\begin{equation}
    A(m)= -0.98\,(m-25)+0.11 \;,\qquad
    B(m)=0.12\,(m-25)+0.17 \; ,
\end{equation}
which yields \cref{eq:bMM} for a magnitude cut  $m=24.5\,$. 

\subsection{Luminosity derivatives}
Derivatives omitted from the analysis but addressed in \cref{sec:lum_derivs}. \\
\textbf{First order $\mathcal{H}/k$:}

\begin{equation}
    \frac{\partial b_{10}}{\partial \ln{L}}\bigg\rvert_{L_c} = \frac{5}{2 \ln{10}} \left[0.98 (1+z) - 0.12 (1+z)^2\right]
\end{equation}
\begin{equation}
    \frac{\partial \ln b_{01}}{\partial \ln{L}}\bigg\rvert_{L_c} = \frac{1}{b_{10}-1}\frac{\partial b_{10}}{\partial \ln{L}}\bigg\rvert_{L_c}
\end{equation}
\textbf{Higher order $\mathcal{H}/k$:}
\begin{equation}
    \frac{\partial\mathcal{Q}}{\partial \ln{L}}\bigg\rvert_{L_c} = -\frac{5}{2 \ln{10}}\frac{\partial\mathcal{Q}}{\partial M}\bigg\rvert_{M_c} =\,\mathcal{Q}\left(1+\alpha-10^{-0.4(M-M_*)} +\mathcal{Q} \right)
\end{equation}

\begin{figure}[h!]
\centering
    \includegraphics[width=.9\textwidth]{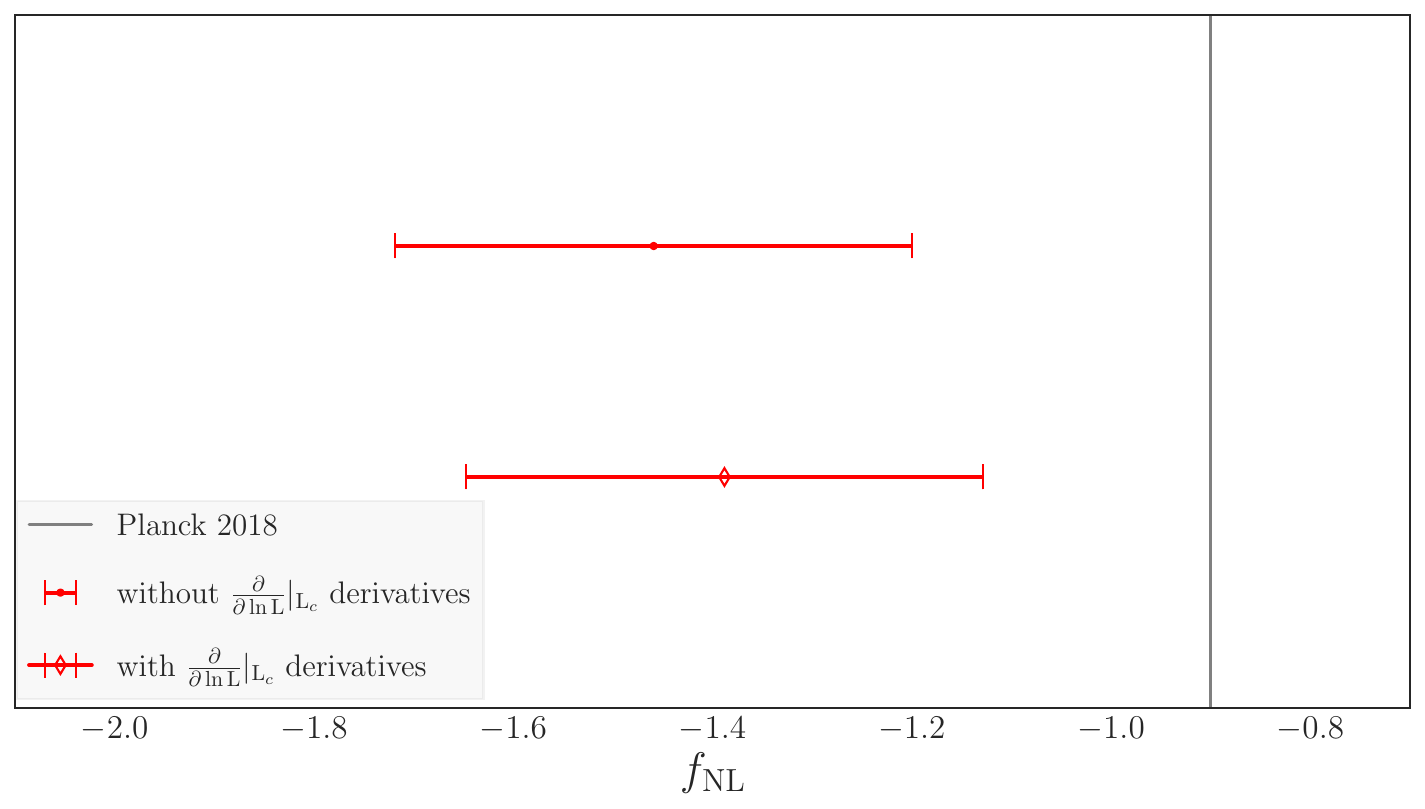}
    \caption{Comparison of the shift on the value of $f_{\mathrm{NL}}$ induced by neglecting GR corrections for MegaMapper without luminosity derivatives (solid line, as shown in \cref{fig:fnl bias comp}) and with luminosity derivatives (diamond).}
    \label{fig:MMbias}
\end{figure}
\newpage
\section{Survey Parameters}\label{app:sp}
\begin{table}[htbp]
\centering
\caption{Survey parameters for \textit{Roman}-like luminosity \textit{Model 1} ($F_{\rm c}=1 \times 10^{-16} \mathrm{erg} \, \mathrm {cm}^{-2} \, \mathrm{s}^{-1}$).\label{tab:Roman_m1}}
\begin{tabular}{ccccc}
\hline
$z$& $V\,[\mathrm{h}^{-3}\,\mathrm{Gpc}^3]$& $n_{\rm g}\,[10^{-3}\,\mathrm{h}^3\,\mathrm{Mpc}^{-3}]$& $\mathcal{E}$&${\cal Q}$\\
\hline\hline
$0.5$ & $0.25$ & $16.15$ & $-2.64$ &$ 0.81$ \\
$0.6$ & $0.32$ & $13.21$ & $-2.82$ &$ 0.90$ \\
$0.7$ & $0.39$ & $11.05$ & $-2.99$ &$ 0.99$ \\
$0.8$ & $0.45$ & $9.39 $& $-3.17$ & $1.08 $\\
$0.9$ & $0.51$ & $8.09 $& $-3.35$ & $1.17 $\\
$1.0$ & $0.56$ & $7.04 $& $-3.52$ & $1.25 $\\
$1.1$ & $0.61$ & $6.19 $& $-3.69$ & $1.34 $\\
$1.2$ & $0.65$ & $5.48 $& $-3.86$ & $ 1.42$\\
$1.3$ & $0.68$ & $4.89 $& $-2.03$ & $1.50 $\\
$1.4$ & $0.71$ & $4.03 $& $-2.19$ & $1.58 $\\
$1.5$ & $0.74$ & $3.35 $& $-2.35$ & $1.66 $\\
$1.6$ & $0.76$ & $2.81 $& $-2.50$ & $1.74 $\\
$1.7$ & $0.78$ & $2.38 $& $-2.65$ & $1.81 $\\
$1.8$ & $0.79$ & $2.02 $& $-2.80$ & $1.89 $\\
$1.9$ & $0.80$ & $1.74 $& $-2.93$ & $1.96 $\\
\hline
\end{tabular}
\end{table}

\begin{table}[htbp]
\centering
\caption{Survey parameters for \textit{Roman}-like luminosity \textit{Model 3} ($F_{\rm c}=1 \times 10^{-16} \mathrm{erg} \, \mathrm {cm}^{-2} \, \mathrm{s}^{-1}$).\label{tab:Roman_m3}}
\begin{tabular}{ccccc}
\hline
$z$& $V\,[\mathrm{h}^{-3}\,\mathrm{Gpc}^3]$& $n_{\rm g}\,[10^{-3}\,\mathrm{h}^3\,\mathrm{Mpc}^{-3}]$& $\mathcal{E}$&$\mathcal{Q}$\\
\hline\hline
$0.5$ & $0.25$ & $9.33$ & $-4.91$ & $1.07 $\\
$0.6$ & $0.32$ & $7.82$ & $-4.71$ & $1.14 $\\
$0.7$ & $0.39$ & $6.65$ & $-4.53$ & $1.21 $\\
$0.8$ & $0.45$ & $5.70$ & $-4.38$ & $1.29 $\\
$0.9$ & $0.51$ & $4.90$ & $-4.24$ & $1.36 $\\
$1.0$ & $0.56$ & $4.22$ & $-4.12$ & $1.44 $\\
$1.1$ & $0.61$ & $3.64$ & $-4.02$ & $1.52 $\\
$1.2$ & $0.65$ & $3.13$ & $-3.92$ & $ 1.60$\\
$1.3$ & $0.68$ & $2.69$ & $-3.84$ & $1.69 $\\
$1.4$ & $0.71$ & $2.31$ & $-3.76$ & $1.77 $\\
$1.5$ & $0.74$ & $1.97$ & $-3.68$ & $1.85 $\\
$1.6$ & $0.76$ & $1.69$ & $-3.60$ & $1.93 $\\
$1.7$ & $0.78$ & $1.44$ & $-3.53$ & $2.01 $\\
$1.8$ & $0.79$ & $1.23$ & $-3.45$ & $2.09 $\\
$1.9$ & $0.80$ & $1.05$ & $-3.37$ & $2.16 $\\
\hline
\end{tabular}
\end{table}

\begin{table}[htbp]
\centering
\caption{DESI BGS survey parameters ($m_{\rm c}=20$).\label{tab:DESI}}
\begin{tabular}{ccccc}
\hline
$z$& $V\,[\mathrm{h}^{-3}\,\mathrm{Gpc}^3]$& $n_{\rm g}\,[10^{-3}\,\mathrm{h}^3\,\mathrm{Mpc}^{-3}]$& $\mathcal{E}$&${\cal Q}$\\
\hline\hline
$0.1$ & $0.11$ & $38.38$ & $-2.66$ &$ 0.55$ \\
$0.2$ & $0.38$ & $15.88$ & $-3.16$ &$ 0.94$ \\
$0.3$ & $0.76$ & $6.42 $& $-3.83$ & $1.50 $\\
$0.4$ & $1.21$ & $2.17 $& $-4.75$ & $2.31 $\\
$0.5$ & $1.69$ & $0.56 $& $-6.03$ & $3.44 $\\
$0.6$ & $2.16$ & $0.10 $& $-7.76$ & $4.94 $\\
\hline
\end{tabular}
\end{table}

\begin{table}[htbp]
\centering
\caption{Survey parameters for \textit{Euclid}-like luminosity \textit{Model 1}, benchmark flux cut ($F_{\rm c}=3 \times 10^{-16} \mathrm{erg} \, \mathrm {cm}^{-2} \, \mathrm{s}^{-1}$).\label{tab:Euclid_m1}}
\begin{tabular}{ccccc}
\hline
$z$& $V\,[\mathrm{h}^{-3}\,\mathrm{Gpc}^3]$& $n_{\rm g}\,[10^{-3}\,\mathrm{h}^3\,\mathrm{Mpc}^{-3}]$& $\mathcal{E}$&${\cal Q}$\\
\hline\hline
$0.7$ & $2.78$ & $2.72$ & $-4.33$ & $1.66 $\\
$0.8$ & $3.23$ & $1.98$ & $-4.76$ & $1.87 $\\
$0.9$ & $3.65$ & $1.46$ & $-5.19$ & $2.09 $\\
$1.0$ & $4.03$ & $1.09$ & $-5.62$ & $2.30 $\\
$1.1$ & $4.36$ & $0.82$ & $-6.05$ & $2.52 $\\
$1.2$ & $4.65$ & $0.62$ & $-6.48$ & $ 2.73$\\
$1.3$ & $4.90$ & $0.48$ & $-4.91$ & $2.94 $\\
$1.4$ & $5.12$ & $0.34$ & $-5.33$ & $3.15 $\\
$1.5$ & $5.30$ & $0.25$ & $-5.74$ & $3.35 $\\
$1.6$ & $5.45$ & $0.18$ & $-6.15$ & $3.56 $\\
$1.7$ & $5.58$ & $0.13$ & $-6.55$ & $3.76 $\\
$1.8$ & $5.68$ & $0.10$ & $-6.94$ & $3.95 $\\
$1.9$ & $5.76$ & $0.07$ & $-7.32$ & $4.14 $\\
$2.0$ & $5.83$ & $0.06$ & $-7.69$ & $ 4.33$\\
\hline
\end{tabular}
\end{table}

\begin{table}[htbp]
\centering
\caption{Survey parameters for \textit{Euclid}-like luminosity \textit{Model 3}, benchmark flux cut ($F_{\rm c}=2 \times 10^{-16} \mathrm{erg} \, \mathrm {cm}^{-2} \, \mathrm{s}^{-1}$).\label{tab:Euclid_m3}}
\begin{tabular}{ccccc}
\hline
$z$& $V\,[\mathrm{h}^{-3}\,\mathrm{Gpc}^3]$& $n_{\rm g}\,[10^{-3}\,\mathrm{h}^3\,\mathrm{Mpc}^{-3}]$& $\mathcal{E}$&${\cal Q}$\\
\hline
$0.9$ & $3.65$ & $1.56$ & $-6.13$ & $1.97$ \\
$1.0$ & $4.03$ & $1.26$ & $-5.93$ & $2.07$ \\
$1.1$ & $4.36$ & $1.02$ & $-5.73$ & $2.17$ \\
$1.2$ & $4.65$ & $0.82$ & $-5.53$ & $2.26$\\
$1.3$ & $4.90$ & $0.67$ & $-5.32$ & $2.34$ \\
$1.4$ & $5.12$ & $0.54$ & $-5.11$ & $2.41$ \\
$1.5$ & $5.30$ & $0.44$ & $-4.91$ & $2.47$ \\
$1.6$ & $5.45$ & $0.36$ & $-4.71$ & $2.52$ \\
$1.7$ & $5.58$ & $0.29$ & $-4.51$ & $2.57$ \\
$1.8$ & $5.68$ & $0.24$ & $-4.32$ & $2.61$ \\
\hline
\end{tabular}
\end{table}

\begin{table}[htbp]
\centering
\caption{MegaMapper LBG survey parameters ($m_{\rm c}=24.5$).\label{tab:MM}}
\begin{tabular}{ccccc}
\hline
$z$& $V\,[\mathrm{h}^{-3}\,\mathrm{Gpc}^3]$& $n_{\rm g}\,[10^{-3}\,\mathrm{h}^3\,\mathrm{Mpc}^{-3}]$& $\mathcal{E}$&${\cal Q}$\\
\hline
$2.1$ & $8.09$ & $2.27$ & $1.80 $& $1.92 $\\
$2.2$ & $8.15$ & $1.81$ & $2.52 $& $1.97 $\\
$2.3$ & $8.20$ & $1.42$ & $2.81 $& $2.02 $\\
$2.4$ & $8.22$ & $1.13$ & $2.50 $& $2.08 $\\
$2.5$ & $8.23$ & $0.92$ & $1.62 $& $2.14 $\\
$2.6$ & $8.23$ & $0.77$ & $0.40 $& $2.21 $\\
$2.7$ & $8.22$ & $0.67$ & $-0.79$ &$ 2.28$ \\
$2.8$ & $8.19$ & $0.60$ & $-1.63$ &$ 2.36$ \\
$2.9$ & $8.16$ & $0.55$ & $-2.01$ &$ 2.45$ \\
$3.0$ & $8.12$ & $0.50$ & $-1.99$ &$ 2.55$ \\
$3.1$ & $8.07$ & $0.46$ & $-1.71$ &$ 2.65$ \\
$3.2$ & $8.02$ & $0.41$ & $-1.29$ &$ 2.76$ \\
$3.3$ & $7.96$ & $0.37$ & $-0.86$ &$ 2.88$ \\
$3.4$ & $7.90$ & $0.32$ & $-0.50$ &$ 3.00$ \\
$3.5$ & $7.83$ & $0.28$ & $-0.23$ &$ 3.12$ \\
$3.6$ & $7.77$ & $0.25$ & $-0.22$ &$ 3.24$ \\
$3.7$ & $7.70$ & $0.21$ & $-0.39$ &$ 3.36$ \\
$3.8$ & $7.62$ & $0.19$ & $-0.77$ &$ 3.46$ \\
$3.9$ & $7.55$ & $0.17$ & $-1.26$ &$ 3.54$ \\
$4.0$ & $7.48$ & $0.15$ & $-1.69$ &$ 3.61$ \\
$4.1$ & $7.40$ & $0.13$ & $-1.99$ &$ 3.65$ \\
$4.2$ & $7.32$ & $0.12$ & $-2.09$ &$ 3.68$ \\
$4.3$ & $7.25$ & $0.11$ & $-1.94$ &$ 3.70$ \\
$4.4$ & $7.17$ & $0.10$ & $-1.47$ &$ 3.71$ \\
$4.5$ & $7.09$ & $0.09$ & $-0.65$ &$ 3.71$ \\
$4.6$ & $7.02$ & $0.08$ & $0.59$ &$ 3.71$ \\
$4.7$ & $6.94$ & $0.07$ & $2.32 $& $3.70 $\\
$4.8$ & $6.86$ & $0.06$ & $4.63 $& $3.70 $\\
$4.9$ & $6.79$ & $0.05$ & $7.64 $& $3.71 $\\
$5.0$ & $6.71$ & $0.03$ & $11.55$ &$ 3.72$ \\
\hline
\end{tabular}
\end{table}
\clearpage



\newpage
\bibliographystyle{JHEP}
\bibliography{biblio}

\end{document}